\documentclass[aps, prd,preprintnumbers,10pt, showpacs, onecolumn,eqsecnum,floatfix,a4paper,nofootinbib,superscriptaddress, reprint]{revtex4-1}

\usepackage{xcolor}
\usepackage{leftidx}
\usepackage{amsmath,amssymb,graphicx}
\usepackage{bm,lipsum}
\usepackage{times}
\usepackage{microtype}
\usepackage[utf8]{inputenc}
\usepackage{booktabs}
\usepackage{subfigure}
\usepackage[normalem]{ulem}
\usepackage[varg]{txfonts}
\usepackage{bm,graphics,graphicx,epsfig,color,ulem}
\usepackage[colorlinks, pdfborder={0 0 0}]{hyperref}
\usepackage{multirow}
\definecolor{LinkColor}{rgb}{0.75, 0, 0}
\definecolor{CiteColor}{rgb}{0, 0.5, 0.5}
\definecolor{UrlColor}{rgb}{0, 0, 0.75}
\hypersetup{linkcolor=LinkColor}
\hypersetup{citecolor=CiteColor}
\hypersetup{urlcolor=UrlColor}
\allowdisplaybreaks
\newcommand{\bigO}{\mathcal{O}}
\newcommand{\xb}{\bar{x}}
\newcommand{\xp}{\tilde{x}}
\newcommand{\eb}{\bar{e_t}}
\newcommand{\ep}{\tilde{e_t}}

\newcommand{\lb}{\bar{l}}
\newcommand{\lp}{\tilde{l}}
\newcommand{\lab}{\bar{\lambda}}
\newcommand{\laP}{\tilde{\lambda}}
\newcommand{\ub}{\bar{u}}

\newcommand{\la}{\langle}
\newcommand{\ra}{\rangle}

\def\no{\nonumber\\}

\newcommand{\mem}{\textnormal{mem}}

\newcommand{\ud}{\mathrm{d}}
\newcommand{\ui}{\mathrm{i}} 
\newcommand{\ue}{\mathrm{e}}

\def\be{\begin{equation}} 
\def\ee{\end{equation}} 
\def\ba{\begin{eqnarray}}
\def\ea{\end{eqnarray}}

\def\la{\langle} \def\ra{\rangle}
\def\bse{\begin{subequations}} 
\def\ese{\end{subequations}}
\def\rl{\right.\nonumber\\&\left.}
\def\rrll{\right.\right.\nonumber\\&\left.\left.}

\begin{document}
%%%%%%%%%%%%%%%%%%%%%%%%%%%%%%%%%%%%%%%%%%%%%%%%%%%%%%%%%%%%%%%%%%%%%%%%%%%%%%%%%%%%%%%%%%%%%% 
%-----------------------------------------TITLE----------------------------------------------%
%%%%%%%%%%%%%%%%%%%%%%%%%%%%%%%%%%%%%%%%%%%%%%%%%%%%%%%%%%%%%%%%%%%%%%%%%%%%%%%%%%%%%%%%%%%%%% 
\title{Linear momentum flux from inspiralling compact binaries in quasi-elliptical orbits at 2.5 Post-Newtonian order}
\author{Shilpa Kastha}\email{shilpa.kastha@aei.mpg.de}
\affiliation{Albert-Einstein-Institut, Max-Planck-Institut f{\"u}r Gravitationsphysik, Callinstra{\ss}e 38, 30167 Hannover, Germany}
\affiliation{Leibniz Universit{\"a}t Hannover, 30167 Hannover, Germany}
%\author{Chandra Kant Mishra} %\email{ckm@iitm.ac.in}
%\affiliation{Indian Institute of Technology, Chennai, India}
%\author{K. G. Arun} %\email{kgarun@cmi.ac.in} 
%\affiliation{Chennai Mathematical Institute, Siruseri, 603103, India}
%:\affiliation{Institute for Gravitation and the Cosmos, Department of Physics, Penn State University, University Park PA 16802, USA}
\date{\today}
%%%%%%%%%%%%%%%%%%%%%%%%%%%%%%%%%%%%%%%%%%%%%%%%%%%%%%%%%%%%%%%%%%%%%%%%%%%%%%%%%%%%%%%%%%%%%% 
%---------------------------------------Abstract----------------------------------------------%
%%%%%%%%%%%%%%%%%%%%%%%%%%%%%%%%%%%%%%%%%%%%%%%%%%%%%%%%%%%%%%%%%%%%%%%%%%%%%%%%%%%%%%%%%%%%%% 
\begin{abstract}
	Emission of anisotropic gravitational radiation from compact binary system leads to a flux of linear momentum. This results in recoil of the system. We investigate the rate of loss of Linear momentum flux in the far zone of the source using various mass type and current type multipole moments for inspiralling compact binary merger in quasi-elliptical orbits at 2.5 Post Newtonian order. We compute the linear momentum flux  accurate upto $\mathcal{O}(e_t)$ in harmonic coordinate. A 2.5 Post Newtonian Quasi-Keplarian representation of the parametric solution to the Post Newtonian equation of motion for the compact binary system has been adopted here. We also provide a closed form expression for the accumulated linear momentum from remote past through the binary evolution.
\end{abstract}
\maketitle

\section{Introduction}
Anisotropic gravitational radiation leads to a flux of linear momentum from the compact binary system~\cite{Peres62,Papa71}. To conserve the total linear momentum, the system recoils. The direction of recoil changes continuously over an orbit. As a result, for a perfectly circular trajectory, no net recoil builds up over an orbit. On the contrary, for inspiralling compact binaries, the recoil accumulates over the inspiralling orbits and imparts a kick to the remnant at the merger. 
A reasonably high kick imparted to the binary black hole (BBH) merger could be of great importance in understanding the structure formation of globular clusters. If the kick is greater than the escape velocity of the host galaxy, the remnant black hole(BH) may even be ejected~\cite{Komossa2008} from the galaxy. Even if the kick is not high enough to eject the remnant BH, it might cause significant dynamical changes at the core of the galaxy. A detailed discussion on various astrophysical aspects of BH kicks can be found in ref~\cite{Merritt2004}.

Though a compact binary merger may have significant eccentricity at the birth, due to the gravitational radiation, it gets circularized~\cite{Peters&Mathews,Peters1964}. By the time their gravitational wave (GW) frequency enters the sensitivity bands of the ground based interferometric GW detectors, they may have negligible eccentricity.
Yet there may be astrophysical processes which may retain their eccentricity even in the late stages of their dynamical evolution. For example, in dense stellar clusters, interactions between pairs of BBH systems may eject one of the BHs leading to the formation of a stable hierarchical triple system. If the two orbital planes are tilted with
respect to each other, the third body can increase the eccentricity of
the inner binary via Kozai mechanism\cite{kozai1962}.
Binaries in such hierarchical triple systems may have non-zero eccentricities even towards the late stages of the inspiral. Further, binary neutron star systems in globular
clusters may have a thermal distribution
of eccentricities~\cite{BenacquistaLiving} if formed by exchange interactions as opposed to the formation scenario through the common envelope. 
Similarly, there have been mechanisms proposed for binaries consisting of supermassive BHs in the LISA band~\cite{Gultekin:2004pm,Hoang:2019kye, Hoang2018} which may have detectable eccentricities. The recent detection~\cite{Abbott:2020tfl, Abbott:2020mjq} of the heaviest BBH merger event to date, GW190521, has been argued to have high eccentricity~\cite{CalderonBustillo:2020odh, Isobel_et_al, gayathri2020gw190521} at merger.
Motivated by these scenarios, we study the emission of linear momentum flux (LMF) in case of non-spinning compact binary systems in quasi-elliptical orbits.

The first formal treatment of gravitational recoil for a general self-gravitating system in linearized gravity is explored in refs.~\cite{Peres62, Bek73}. It is valid for any kind of motion (rotational, vibrational or any other kind) given the source is localized within a finite volume. Later, within the Post-Newtonian (PN) framework, the leading order contribution (Newtonian) to the LMF and recoil of a compact binary system is discussed in refs.~\cite{Fitchett83, Fitchett84}. 
The first PN correction to this was computed by Wiseman~\cite{Wi92} and the quasi-circular orbit scenario was discussed as a special case there. Much later, a closed form expression for the recoil in case of compact binary in quasi-circular orbit is quoted in~\cite{BQW05} at the second post-Newtonian (PN) order. Its extension to 2.5PN order, accounting for the radiation reaction effects, is discussed in ref.\cite{MAI12}. According to these studies, the BH recoil for nonspinning systems could be in the range 74-250 km $\rm{s}^{-1}$. Using the effective one-body(EOB) approach, the recoil estimates for BBH is obtained considering the contributions from inspiral, plunge and ringdown phases in ref.~\cite{Gopu2006}. The typical estimate obtained here, lies in the range $49$-$172$ km $\rm{s}^{-1}$.
In ref.~\cite{Favata2004}, BH perturbation theory is used to compute the accumulated recoil up to the innermost stable circular orbit (ISCO) ($10$-$100$ km $\rm{s}^{-1}$) for a system where a test particle inspirals into a BH. In principle, these estimates are valid for extreme mass ratio inspirals but they have extrapolated the results till mass ratio of about $\sim0.4$.

Along with the various analytical and semi-analytical studies, the recent progress in numerical relativity techniques has led to more accurate estimates for the recoil of the remnant BH. As quoted in refs.~\cite{Campanelli:2004zw, Baker:2006vn,Herrmann_2007, Gonzalez2008}, the recoil velocity can reach up to a few hundreds of km $\rm{s}^{-1}$ while the component masses are nonspinning for compact binaries in quasi-circular orbit. But for the spinning case~\cite{Herrmann:2007ac, Koppitz:2007ev, Gonzalez:2007hi}, the recoil velocity can be of the order of few thousands km $\rm{s}^{-1}$. In case of maximally spinning BHs, it could be as high as 4000 km $\rm{s}^{-1}$~\cite{Campanelli:2007ew}. Such a large recoil velocity may lead to ejection of the merged binary from its host galaxy. A detailed study on multipolar analysis of the gravitational recoil is also discussed in ref.~\cite{Schnittman:2007ij}. They have explored the build up of the recoil through the different phases of the binary evolution, (inspiral+merger+ringdown) due to the relative amplitude and the phases of various modes of the GWs.

Using a formal approach within the Multipolar Post-Minkowskian Post-Newtonian (MPM-PN) formalism, the leading order Newtonian contribution to the LMF and the associated recoil of a compact binary system is explored by Fitchett in refs.~\cite{Fitchett83, Fitchett84}. The authors have assumed the inspiralling binary to be composed of two point particles moving in a Keplerian orbit. A rough estimate of maximum BH recoil quoted in these studies are $\sim 1500 \rm{ km/s}$. Assuming the periastron advance to be small, a crude estimate of the recoil at 1PN is also quoted. The first extension of these estimates at 1PN for binaries moving in generic orbits is explored by Wiseman~\cite{Wi92} and concluded that higher order correction reduces the net momentum ejection.  As a special case, they also studied BNS systems moving in quasi-circular orbits. They found the upper bound on the velocity of the center of mass very near to the coalescence to be 1 km $\rm{s}^{-1}$. 
In another study~\cite{Sopuerta_2007}, the authors showed a 10\% increase in the recoil estimate compared to the quasi-circular case for small eccentricities $(e<0.1)$ using close limit approximation. They also claimed that the maximum recoil takes place for the value of the symmetric mass ratio of around $\eta\sim0.19$ and the magnitude could be as high as $216-242$ km/s.

In order to obtain a correct recoil estimates one needs to compute the linear momentum from the system.
Here we investigate the linear momentum flux at 2.5 post-Newtonian order for compact binaries moving in quasi-elliptical orbits with non-spinning component masses.

The paper is organized as follows. In section~\ref{MultipolarLMF} we discuss the multipolar decomposition of LMF in terms of the various source type multipole moments and their non-linear interactions. In section~\ref{orbital_dynamics_LMF}, we discuss the orbital dynamics and  in section~\ref{inst_r_rdot_v_LMF} we quote the instantaneous contribution to the LMF in terms of the orbital parameters. In section~\ref{KR} and ~\ref{LMF_inst_et_phi} we summarize the keplerian and generalized quasi-keplerian representations (QKR) of the orbital dynamics and re-express the instanteneous contribution to the LMF in terms of the QK parameters. Section~\ref{LMF_herd_et_phi} and ~\ref{post-adiabatic} consists of the computation of  hereditary and the post-adiabatic contributions to the LMF. We quote the complete LMF at 2.5PN in section~\ref{totalLMF}. Finally in section~\ref{accumulated} we compute the accumulated linear momentum through the inspiralling orbits from remote past.

%%%%%%%%%%%%%%%%%%%%%%%%%%%%%%%%%%%%%%%%%%%%%%%%%%%%%%%%%%%%%%%%%%%%%%%%%%%%%%%%
\section{Multipole decomposition of Linear momentum flux}\label{MultipolarLMF}
%%%%%%%%%%%%%%%%%%%%%%%%%%%%%%%%%%%%%%%%%%%%%%%%%%%%%%%%%%%%%%%%%%%%%%%%%%%%%%%%

For an isolated source, gravitational wave generation is well studied under the framework of multipolar decomposition~\cite{Thorne80}. Following~\cite{Thorne80}, we explicitly write down the multipolar decomposition of far-zone linear momentum flux (LMF) for an isolated source at the 2.5 post-Newtonian order in terms  symmetric trace-free (STF) radiative mass type and current type multipole moments [see Eq.(2.1) of ref.~\cite{MAI12}].
%%%%%%%%%%%%%%%%%%%%%%%%%%%%%%%%%%%%%%%%%%%%%%%%%%%%%%%%%%%%%%%%%%%%%%%%%%%%%%%%
%-------------------------Total LMF expression----------------------------- %
%%%%%%%%%%%%%%%%%%%%%%%%%%%%%%%%%%%%%%%%%%%%%%%%%%%%%%%%%%%%%%%%%%%%%%%%%%%%%%%%
\begin{widetext}\begin{eqnarray}
	\label{LMFUV}
	{\mathcal{F}_{i}}&=& \frac{G}{c^7}\,\biggl\{\left[\frac{2}{63}\,
	U^{(1)}_{ijk}\,U^{(1)}_{jk}+\frac{16}{45}\,\varepsilon_{ijk}
	U^{(1)}_{ja}\,V^{(1)}_{ka}\right]
	+{1\over c^2}\left[\frac{1}{1134}\,U^{(1)}_{ijkl}\,U^{(1)}_{jkl}
	+\frac{1}{126}\,\varepsilon_{ijk}
	U^{(1)}_{jab}\,V^{(1)}_{kab}
	+\frac{4}{63}\,V^{(1)}_{ijk}\,V^{(1)}_{jk}\right]\nonumber\\&&
	+{1\over
		c^4}\left[\frac{1}{59400}\,U^{(1)}_{ijklm}\,U^{(1)}_{jklm}+\frac{2}{14175}\,\varepsilon_{ijk}
	U^{(1)}_{jabc}\,V^{(1)}_{kabc}+
	\frac{2}{945}\,V^{(1)}_{ijkl}\,V^{(1)}_{jkl}\right]+{\cal O}\left({1\over c^6}\right) \biggr\}.
\end{eqnarray}\end{widetext}
%%%%%%%%%%%%%%%%%%%%%%%%%%%%%%%%%%%%%%%%%%%%%%%%%%%%%%%%%%%%%%%%%%%%%%%%%%%%%%%%
%%%%%%%%%%%%%%%%%%%%%%%%%%%%%%%%%%%%%%%%%%%%%%%%%%%%%%%%%%%%%%%%%%%%%%%%%%%%%%%%
%
%
Here $U_{K}^{(p)}$ and $V_{K}^{(p)}$ \Big( $K=i_1i_2\cdots i_k$
represents the multi-index structure of the tensors of order $k$ in three dimension\Big) are the $p^{th}$-time derivative of mass-type and current-type radiative multipole moments respectively. $\varepsilon_{ijk}$ is the usual three dimensional Levi-Civita tensor, with a value $+1$ in case of all even permutations and $-1$ for all the odd ones. The multipole moments in the formula, are functions of retarded time $(t-({\it r}/c))$ in radiative coordinates, where $\it r$ and $t$ denote the distance of the source from the observer and the time of observation respectively.

Using the multipolar post Minkowskian (MPM) formalism~\cite{Th80,BD84,BD86,B87,BD88,BD89,BD92,B95, BDI95, BIJ02,DJSdim,BDEI04} the two types of radiative moments, $(U_L, V_L)$ can be expressed in terms of two canonical moments $(M_L,S_L)$ and eventually as a function of all the source multipole moments $(I_L, J_L, X_L, W_L, Y_L, Z_L)$~\cite{BFIS08} at the 2.5PN order.
Every radiative moment has two types of contributions. One of them is only a function of retarded time and hence called the {\it instantaneous} part. The other one depends on the dynamical behavior of the system throughout its entire past and referred to as the {\it hereditary} contributions. These contributions contain information about various multipolar interactions the gravitational wave undergoes, as it propagates from the source to the detector.

For the reader's convenience, here we explicitly quote all the radiative moments in terms of the source moments, accurate up to the order necessary for the present calculation(see Eqs. (3.1)-(3.18b) of refs.~\cite{Mai2015}).
Since the leading order term in the LMF expression (see Eq.~\ref{LMFUV}) consists of the mass quadrupole moment $(U_{ij})$, the desired accuracy of $U_{ij}$ is 2.5PN. Furthermore, the decomposition of mass quadrupole moment into instantaneous and hereditary parts is as follows, 
%%%%%%%%%%%%%%%%%%%%%%%%%%%%%%%%%%%%%%%%%%%%%%%%%%%%%%%%%%%%%%%%%%%%%%%%%%%%%%%%
%--------------------------------------------U2--------------------------------------------- %
%%%%%%%%%%%%%%%%%%%%%%%%%%%%%%%%%%%%%%%%%%%%%%%%%%%%%%%%%%%%%%%%%%%%%%%%%%%%%%%%
\be U_{ij}=U_{ij}^{\rm inst}+U_{ij}^{\rm hered}, \label{Uij} \ee
where the instantaneous and the hereditary parts explicitly read
\begin{widetext} \bse\label{ellWF:U2} \begin{align}\label{ellWF:U2inst}
	U_{ij}^{\rm inst}(U) &= I^{(2)}_{ij} (U) +{G \over c^5}\left\{{1
		\over7}I^{(5)}_{a\langle i}I_{j\rangle a} - {5  \over7} I^{(4)}_{a\langle
		i}I^{(1)}_{j\rangle a} -{2  \over7} I^{(3)}_{a\langle i}I^{(2)}_{j\rangle a}
	+{1  \over3}\varepsilon_{ab\langle i}I^{(4)}_{j\rangle a}J_{b} +4\left[W^{(4)}
	I_{ij}+W^{(3)} I_{ij}^{(1)}\rrll-W^{(2)} I_{ij}^{(2)}- W^{(1)}
	I_{ij}^{(3)}\right]\right\} +\,\, \mathcal{O}\left(\frac{1}{c^7}\right),\\
	%%%%%%%%%%%%%%%%%%%%%%%%%%%%%%%%%%%%%%%%%%%%%%%%%%%%%%%%%%%%%%%%%%%%%%%%%%%%%%%%
	U_{ij}^{\rm hered}(U) &= {2G M \over c^3} \int_{-\infty}^{U} d \tau \left[ \ln
	\left({U-\tau \over 2\tau_0}\right)+{11 \over12} \right] I^{(4)}_{ij} (\tau)
	+{G \over c^5}\left\{-{2 \over7}\int_{-\infty}^{U} d\tau I^{(3)}_{a\langle
		i}(\tau)I^{(3)}_{j\rangle a}(\tau) \right\}\nonumber\\ & + 2\left({G M \over
		c^3}\right)^2\int_{-\infty}^{U} d \tau \left[ \ln^2 \left({U-\tau
		\over2\tau_0}\right)+{57 \over70} \ln\left({U-\tau \over
		2\tau_0}\right)+{124627 \over44100} \right] I^{(5)}_{ij} (\tau) +\,\,
	\mathcal{O}\left(\frac{1}{c^7}\right)\,.  \label{ellWF:U2hered}\end{align} \ese
\end{widetext}
%%%%%%%%%%%%%%%%%%%%%%%%%%%%%%%%%%%%%%%%%%%%%%%%%%%%%%%%%%%%%%%%%%%%%%%%%%%%%%%%
%%%%%%%%%%%%%%%%%%%%%%%%%%%%%%%%%%%%%%%%%%%%%%%%%%%%%%%%%%%%%%%%%%%%%%%%%%%%%%%%

In the above expression, $M$ represents the Arnowitt, Deser and Misner (ADM) mass of the source and hence undergoes relativistic corrections given by $M=m(1-\eta x/2)$. The constant $\tau_0$ is related to an arbitrary length scale $r_0$
by $\tau_0=r_0/c$ and was originally introduced in the MPM formalism.

The required accuracy of mass octupole moment is 1.5PN which is,
%%%%%%%%%%%%%%%%%%%%%%%%%%%%%%%%%%%%%%%%%%%%%%%%%%%%%%%%%%%%%%%%%%%%%%%%%%%%%%%%
%-------------------------------------U3-------------------------------------- %
%%%%%%%%%%%%%%%%%%%%%%%%%%%%%%%%%%%%%%%%%%%%%%%%%%%%%%%%%%%%%%%%%%%%%%%%%%%%%%%%
\begin{align}\label{Uijk} 
	U_{ijk}&=U_{ijk}^{\rm inst}+U_{ijk}^{\rm hered},
\end{align}
and both the parts separately read
\begin{widetext} \begin{subequations}\label{ellWF:U3}\begin{align} U_{ijk}^{\rm
			inst} (U) &= I^{(3)}_{ijk} (U) +{G \over c^5}\left\{ -{4
			\over3}I^{(3)}_{a\langle i}I^{(3)}_{jk\rangle a}-{9 \over4}I^{(4)}_{a\langle
			i}I^{(2)}_{jk\rangle a} + {1 \over4}I^{(2)}_{a\langle i}I^{(4)}_{jk\rangle a} -
		{3 \over4}I^{(5)}_{a\langle i}I^{(1)}_{jk\rangle a} +{1
			\over4}I^{(1)}_{a\langle i}I^{(5)}_{jk\rangle a}+ {1 \over12}I^{(6)}_{a\langle
			i}I_{jk\rangle a}\rl+{1 \over4}I_{a\langle i}I^{(6)}_{jk\rangle a} + {1
			\over5}\varepsilon_{ab\langle i}\left[-12J^{(2)}_{ja}I^{(3)}_{k\rangle
			b}-8I^{(2)}_{ja}J^{(3)}_{k\rangle b} -3J^{(1)}_{ja}I^{(4)}_{k\rangle
			b}-27I^{(1)}_{ja}J^{(4)}_{k\rangle b}-J_{ja}I^{(5)}_{k\rangle
			b}-9I_{ja}J^{(5)}_{k\rangle b} \rrll-{9 \over4}J_{a}I^{(5)}_{jk\rangle
			b}\right]+{12 \over5}J_{\langle i}J^{(4)}_{jk\rangle}+4\left[W^{(2)}
		I_{ijk}-W^{(1)} I_{ijk}^{(1)} +3\, I_{\langle
			ij}Y_{k\rangle}^{(1)}\right]^{(3)}\right\}
		+\,\mathcal{O}\left(\frac{1}{c^6}\right) \label{ellWF:U3inst},\\
		%%%%%%%%%%%%%%%%%%%%%%%%%%%%%%%%%%%%%%%%%%%%%%%%%%%%%%%%%%%%%%%%%%%%%%%%%%%% 
		U_{ijk}^{\rm hered} (U) &={2G M \over c^3} \int_{-\infty}^{U} d\tau\left[ \ln
		\left({U-\tau \over 2\tau_0}\right)+{97 \over60} \right] I^{(5)}_{ijk} (\tau)
		+{G \over c^5}\left\{ \int_{-\infty}^{U} d\tau \left[-{1
			\over3}I^{(3)}_{a\langle i} (\tau)I^{(4)}_{jk\rangle a} (\tau)\rrll-{4
			\over5}\varepsilon_{ab\langle i} I^{(3)}_{ja} (\tau)J^{(3)}_{k\rangle b}
		(\tau)\right]\right\} +\,\mathcal{O}\left(\frac{1}{c^6}\right)
		\label{ellWF:U3hered}.  \end{align}\end{subequations}\end{widetext}
%%%%%%%%%%%%%%%%%%%%%%%%%%%%%%%%%%%%%%%%%%%%%%%%%%%%%%%%%%%%%%%%%%%%%%%%%%%%%%%%
%%%%%%%%%%%%%%%%%%%%%%%%%%%%%%%%%%%%%%%%%%%%%%%%%%%%%%%%%%%%%%%%%%%%%%%%%%%%%%%%

As the other two mass type mulitpole moments $U_{ijkl}$ and $U_{ijklm}$ appear in the LMF at 1PN and 2PN respectively, the desired accuracy for these two are 1.5 PN and 0.5PN respectively, which read, 
%%%%%%%%%%%%%%%%%%%%%%%%%%%%%%%%%%%%%%%%%%%%%%%%%%%%%%%%%%%%%%%%%%%%%%%%%%%%%%%%
%--------------------------------------------U4--------------------------------------------- %
%%%%%%%%%%%%%%%%%%%%%%%%%%%%%%%%%%%%%%%%%%%%%%%%%%%%%%%%%%%%%%%%%%%%%%%%%%%%%%%%	
\begin{align}\label{Uijkl} 
U_{ijkl}&=U_{ijkl}^{\rm inst}+U_{ijkl}^{\rm hered},
\end{align}

\begin{widetext}\begin{subequations}\label{ellWF:U4}\begin{align} U_{ijkl}^{\rm
			inst}(U) &= I^{(4)}_{ijkl} (U) + {G \over c^3} \left\{ -{21
			\over5}I^{(5)}_{\langle ij}I_{kl\rangle }
		- {63 \over5}I^{(4)}_{\langle ij}I^{(1)}_{kl\rangle }- {102
			\over5}I^{(3)}_{\langle ij}I^{(2)}_{kl\rangle }\right\}
		+\,\mathcal{O}\left(\frac{1}{c^5}\right)\label{ellWF:U4inst},\\
		U_{ijkl}^{\rm hered} (U) &={G \over c^3} \left\{ 2 M \int_{-\infty}^{U} d \tau
		\left[ \ln \left({U-\tau \over 2\tau_0}\right)+{59 \over30} \right]
		I^{(6)}_{ijkl}(\tau)  +{2 \over5}\int_{-\infty}^{U} d\tau I^{(3)}_{\langle
			ij}(\tau)I^{(3)}_{kl\rangle }(\tau) \right\}
		+\,\mathcal{O}\left(\frac{1}{c^5}\right).\label{ellWF:U4hered}
		\end{align}\end{subequations}\end{widetext}
%%%%%%%%%%%%%%%%%%%%%%%%%%%%%%%%%%%%%%%%%%%%%%%%%%%%%%%%%%%%%%%%%%%%%%%%%%%%%%%%
%%%%%%%%%%%%%%%%%%%%%%%%%%%%%%%%%%%%%%%%%%%%%%%%%%%%%%%%%%%%%%%%%%%%%%%%%%%%%%%%
%%%%%%%%%%%%%%%%%%%%%%%%%%%%%%%%%%%%%%%%%%%%%%%%%%%%%%%%%%%%%%%%%%%%%%%%%%%%%%%%
%--------------------------------------------U5---------------------------------------------
%%%%%%%%%%%%%%%%%%%%%%%%%%%%%%%%%%%%%%%%%%%%%%%%%%%%%%%%%%%%%%%%%%%%%%%%%%%%%%%%
\begin{align} U_{ijklm}^{\rm
		inst}(T_R) &= I^{(4)}_{ijklm} (T_R) +\,\mathcal{O}\left(\frac{1}{c^4}\right).\label{U5inst}
\end{align}
%%%%%%%%%%%%%%%%%%%%%%%%%%%%%%%%%%%%%%%%%%%%%%%%%%%%%%%%%%%%%%%%%%%%%%%%%%%%%%%%
%%%%%%%%%%%%%%%%%%%%%%%%%%%%%%%%%%%%%%%%%%%%%%%%%%%%%%%%%%%%%%%%%%%%%%%%%%%%%%%%
Among the current type moments, current quadrupole moment is needed to be evaluated at 2.5PN order. 
%------------------------------------------- %---------------------------------------------- %
%--------------------------------------------V2--------------------------------------------- %
%------------------------------------------- %---------------------------------------------- %	
\begin{align} V_{ij}&=V_{ij}^{\rm inst}+V_{ij}^{\rm hered},
	\label{Vij} \end{align}
The instantaneous and the hereditary parts of the current quadrupole moments read
\begin{widetext}\begin{subequations}\label{ellWF:V2}\begin{align} V_{ij}^{\rm
			inst} (U) &= J^{(2)}_{ij} (U) + {G \over7\,c^{5}}\left\{4J^{(2)}_{a\langle
			i}I^{(3)}_{j\rangle a}+8I^{(2)}_{a\langle i}J^{(3)}_{j\rangle a}
		+17J^{(1)}_{a\langle i}I^{(4)}_{j\rangle a}-3I^{(1)}_{a\langle
			i}J^{(4)}_{j\rangle a}+9I_{a\langle i}I^{(5)}_{j\rangle a} -3I_{a\langle
			i}J^{(5)}_{j\rangle a}\rl-{1 \over4}J_{a}I^{(5)}_{ija} -7\varepsilon_{ab\langle
			i}J_{a}J^{(4)}_{j\rangle b} +{1 \over2}\varepsilon_{ac\langle
			i}\left[3I^{(3)}_{ab}I^{(3)}_{j\rangle bc} +{353 \over24}I^{(2)}_{j\rangle
			bc}I^{(4)}_{ab} -{5 \over12}I^{(2)}_{ab}I^{(4)}_{j\rangle bc}+{113
			\over8}I^{(1)}_{j\rangle bc}I^{(5)}_{ab} \rrll-{3
			\over8}I^{(1)}_{ab}I^{(5)}_{j\rangle bc}+{15 \over4}I_{j\rangle bc}I^{(6)}_{ab}
		+{3 \over8}I_{ab}I^{(6)}_{j\rangle bc}\right]+14\,\left[\varepsilon_{ab\langle
			i}\left(-I_{j\rangle b}^{(3)}W_{a}-2I_{j\rangle b}Y_{a}^{(2)} +I_{j\rangle
			b}^{(1)}Y_{a}^{(1)}\right)\rrll+3J_{\langle i}Y_{j\rangle
		}^{(1)}-2J_{ij}^{(1)}W^{(1)}\right]^{(2)} \right\}
		+\,\mathcal{O}\left(\frac{1}{c^6}\right)\,,\label{ellWF:V2inst}\\
		%%%%%%%%%%%%%%%%%%%%%%%%%%%%%%%%%%%%%%%%%%%%%%%%%%%%%%%%%%%%%%%%%%%%%%%%%%%%%%%%%%%%%%%%%
		V_{ij}^{\rm hered} (U) &= {2G M \over c^3} \int_{-\infty}^{U} d \tau \left[ \ln
		\left({U-\tau \over 2\tau_0}\right)+{7 \over6} \right] J^{(4)}_{ij}(\tau)
		+\,\mathcal{O}\left(\frac{1}{c^6}\right)\,.\label{ellWF:V2hered}
		\end{align}\end{subequations}\end{widetext}
%------------------------------------------- %---------------------------------------------- %
%------------------------------------------- %---------------------------------------------- %
For current octupole moment the two contributions are the following 
%------------------------------------------- %---------------------------------------------- %
%--------------------------------------------V3--------------------------------------------- %
%------------------------------------------- %---------------------------------------------- %	
\begin{align} V_{ijk}&=V_{ijk}^{\rm
	inst}+V_{ijk}^{\rm hered}, \label{ellWF:V3decom} \end{align} 
where
$V_{ijk}^{\rm inst}$ and $V_{ijk}^{\rm hered}$ are given in terms of the source multipole
moments as  
\begin{widetext}\begin{subequations}\label{ellWF:V3}\begin{align}
		V_{ijk}^{\rm inst} (U) &= J^{(3)}_{ijk} (U) + {G \over c^3} \left\{ {1
			\over10}\varepsilon_{ab\langle i}I^{(5)}_{ja}I_{k\rangle b}- {1
			\over2}\varepsilon_{ab\langle i}I^{(4)}_{ja}I^{(1)}_{k\rangle b} - 2 J_{\langle
			i}I^{(4)}_{jk\rangle } \right\}
		+\,\mathcal{O}\left(\frac{1}{c^5}\right),\label{ellWF:V3inst}\\ V_{ijk}^{\rm
			hered} (U) &= {2 G M \over c^3}\int_{-\infty}^{U} d \tau \left[ \ln
		\left({U-\tau \over 2\tau_0}\right)+{5 \over3} \right] J^{(5)}_{ijk} (\tau)
		+\,\mathcal{O}\left(\frac{1}{c^5}\right).\label{ellWF:V3hered}
		\end{align}\end{subequations}\end{widetext}
%%%%%%%%%%%%%%%%%%%%%%%%%%%%%%%%%%%%%%%%%%%%%%%%%%%%%%%%%%%%%%%%%%%%%%%%%%%%%%%%%%%%%%%%%
For one other current type moments $V_{ijkl}$, we only need the instantaneous part to obtain LMF at 2.5PN, which reads,
%------------------------------------------- %---------------------------------------------- %	
\begin{align} V_{ijkl}^{\rm
		inst}(T_R) &=J^{(4)}_{ijkl}(T_R) 
	+\mathcal{O}\left(\frac{1}{c^3}\right).\label{V4inst}
\end{align}
%%%%%%%%%%%%%%%%%%%%%%%%%%%%%%%%%%%%%%%%%%%%%%%%%%%%%%%%%%%%%%%%%%%%%%%%%%%%%%%%%%%%%
Using Eqs.~\eqref{Uij}-\eqref{V4inst} we obtain the closed form expression for LMF at 2.5PN in terms of the source multipole moments. Similar to the radiative moments, the LMF also admits a decomposition into instantaneous and hereditary parts.
The instantaneous and the hereditary parts indicate two distinct physical processes and their evaluations need separate treatments. Thus, for our convenience, we explicitly write the two types of contributions ( instantaneous and hereditary) to linear momentum flux separately as follows,
%------------------------------------------- %---------------------------------------------- %
%-------------------------------LMF instantaneous terms----------------------------------- %
%------------------------------------------- %---------------------------------------------- %	
\begin{align}
	{\mathcal{F}_{i}}=\left({\mathcal{F}_{i}}\right)_{\rm inst}+\left({\mathcal{F}_{i}}\right)_{\rm hered},
\end{align}
where the instantaneous part up to 2.5PN in terms of the source multipole moments is (see Eq.(2.2) of ref.~\cite{MAI12})
\begin{widetext}
\begin{eqnarray}
\label{eq:LMF-inst-IJ}
\left({\mathcal{F}_{i}}\right)_{\rm inst}&=&
\frac{G}{c^7}\,\left\{\frac{2}{63}\,
I^{(4)}_{ijk}\,I^{(3)}_{jk}+\frac{16}{45}\,\varepsilon_{ijk}
I^{(3)}_{ja}\,J^{(3)}_{ka}
+{1\over c^2}\left[\frac{1}{1134}\,I^{(5)}_{ijkl}\,I^{(4)}_{jkl}+
\frac{4}{63}\,J^{(4)}_{ijk}\,J^{(3)}_{jk}+\frac{1}{126}\,\varepsilon_{ijk}
I^{(4)}_{jab}\,J^{(4)}_{kab}\right]
\right.\nonumber\\&&\left.
+{1\over c^4}\left[\frac{1}{59400}\,I^{(6)}_{ijklm}\,I^{(5)}_{jklm}+
\frac{2}{945}\,J^{(5)}_{ijkl}\,J^{(4)}_{jkl}+\frac{2}{14175}\,\varepsilon_{ijk}
I^{(5)}_{jabc}\,J^{(5)}_{kabc}\right]
\right.\nonumber\\&&\left.
+{G\over c^5}\left[{2\over63}\left(I_{ijk}^{(4)}\left[{1\over7}\,I_{a\la j}^{(6)}I_{k\ra a}
-{4\over7}\,I_{a\la j}^{(5)}I_{k\ra a}^{(1)}
-I_{a\la j}^{(4)}I_{k\ra a}^{(2)}
-{4\over7}\,I_{a\la j}^{(3)}I_{k\ra a}^{(3)}
+{1\over3}\,\varepsilon_{ab\la j}I_{k\ra a}^{(5)}J_{b}
\right.\right.\right.\right.\nonumber\\&&\left.\left.\left.\left.
+4\left[W^{(2)}I_{jk}-W^{(1)}I_{jk}^{(1)}\right]^{(3)}\right]
+I_{jk}^{(3)}\left[
-{43\over12}\,I_{a\la i}^{(4)}I_{jk\ra a}^{(3)}
-{17\over12}\,I_{a\la i}^{(3)}I_{jk\ra a}^{(4)}
-3\,I_{a\la i}^{(5)}I_{jk\ra a}^{(2)}
\right.\right.\right.\right.\nonumber\\&&\left.\left.\left.\left.
+{1\over2}\,I_{a\la i}^{(2)}I_{jk\ra a}^{(5)}
-{2\over3}\,I_{a\la i}^{(6)}I_{jk\ra a}^{(1)}
+{1\over2}\,I_{a\la i}^{(1)}I_{jk\ra a}^{(6)}
+{1\over12}\,I_{a\la i}^{(7)}I_{jk\ra a}
+{1\over4}\,I_{a\la i}I_{jk\ra a}^{(7)}
+{1\over5}\,\varepsilon_{ab\la i}\left(
-12\,J_{j\ud{a}}^{(3)}I_{k\ra b}^{(3)}
\right.\right.\right.\right.\right.\nonumber\\&&\left.\left.\left.\left.\left.
-12\,I_{j\ud{a}}^{(3)}J_{k\ra b}^{(3)}
-15\,J_{j\ud{a}}^{(2)}I_{k\ra b}^{(4)}
-35\,I_{j\ud{a}}^{(2)}J_{k\ra b}^{(4)}
-4\,J_{j\ud{a}}^{(1)}I_{k\ra b}^{(5)}-36\,I_{j\ud{a}}^{(1)}J_{k\ra b}^{(5)}
-J_{j\ud{a}}I_{k\ra b}^{(6)}
-9\,I_{j\ud{a}}J_{k\ra b}^{(6)}
\right.\right.\right.\right.\right.\nonumber\\&&\left.\left.\left.\left.\left.
-{9\over4}\,J_{\ud{a}}I_{jk\ra b}^{(6)}
\right)
+{12\over5}\,J_{\la i}J_{jk\ra}^{(5)}
+4\,\left[W^{(2)}I_{ijk}-W^{(1)}I_{ijk}^{(1)}+3\,I_{\la ij}Y_{k\ra}^{(1)}\right]^{(4)}\right]\right)
+{16\over45}\,\varepsilon_{ijk}\left(I_{jp}^{(3)}
\right.\right.\right.\nonumber\\&&\left.\left.\left.
\left[{4\over7}\,J_{a\la k}^{(3)}I_{p\ra a}^{(3)}
+{8\over7}\,I_{a\la k}^{(3)}J_{p\ra a}^{(3)}
+3\,J_{a\la k}^{(2)}I_{p\ra a}^{(4)}+{5\over7}\,I_{a\la k}^{(2)}J_{p\ra a}^{(4)}
+{26\over7}\,J_{a\la k}^{(1)}I_{p\ra a}^{(5)}-{6\over7}\,I_{a\la k}^{(1)}J_{p\ra a}^{(5)}
+{9\over7}\,J_{a\la k}I_{p\ra a}^{(6)}
\right.\right.\right.\right.\nonumber\\&&\left.\left.\left.\left.
-{3\over7}\,I_{a\la k}J_{p\ra a}^{(6)}
-{1\over28}\,J_{a}I_{kpa}^{(6)}
-\varepsilon_{ab\la k}J_{\ud{a}}J_{p\ra b}^{(5)}
+{1\over14}\,\varepsilon_{ac\la k}
\left({425\over24}\,I_{p\ra bc}^{(3)}I_{ab}^{(4)}+{31\over12}\,I_{p\ra bc}^{(4)}I_{ab}^{(3)}
+{173\over6}\,I_{p\ra bc}^{(2)}I_{ab}^{(5)}
\right.\right.\right.\right.\right.\nonumber\\&&\left.\left.\left.\left.\left.
-{19\over24}\,I_{p\ra bc}^{(5)}I_{ab}^{(2)}
+{143\over8}\,I_{p\ra bc}^{(1)}I_{ab}^{(6)}
+{15\over4}\,I_{p\ra bc}I_{ab}^{(7)}+{3\over8}\,I_{p\ra bc}^{(7)}I_{ab}\right)
+2\left[\varepsilon_{ab\la k}\left(-I_{p\ra b}^{(3)}W_{a}-2\,I_{p\ra b}Y_{a}^{(2)}
\right.\right.\right.\right.\right.\right.\nonumber\\&&\left.\left.\left.\left.\left.\left.
+I_{p\ra b}^{(1)}Y_{a}^{(1)}\right)
+3\,J_{\la k}Y_{p\ra}^{(1)}-2\,J_{kp}^{(1)}W^{(1)}\right]^{(3)}\right]
+J_{kp}^{(3)}\left[{1\over7}\,I_{a\la j}^{(6)}I_{p\ra a}
-{4\over7}\,I_{a\la j}^{(5)}I_{p\ra a}^{(1)}
-I_{a\la j}^{(4)}I_{p\ra a}^{(2)}
-{4\over7}\,I_{a\la j}^{(3)}I_{p\ra a}^{(3)}
\right.\right.\right.\right.\nonumber\\&&\left.\left.\left.\left.
+{1\over3}\,\varepsilon_{ab\la j}I_{p\ra a}^{(5)}J_{b}
+4\left[W^{(2)}I_{jp}-W^{(1)}I_{jp}^{(1)}\right]^{(3)}\right]\right)
+{1\over1134}\,I_{jkl}^{(4)}\left(-20\,I_{\la ij}^{(3)}I_{kl\ra}^{(3)}
-{84\over5}\,I_{\la ij}^{(5)}I_{kl\ra}^{(1)}
-33\,I_{\la ij}^{(4)}I_{kl\ra}^{(2)}
\right.\right.\right.\nonumber\\&&\left.\left.\left.
-{21\over5}\,I_{\la ij}^{(6)}I_{kl\ra}
\right)
+{1\over126}\,\varepsilon_{ijk}\,I_{jpq}^{(4)}\left(
-{2\over5}\,\varepsilon_{ab\la k}I_{p\ud{a}}^{(5)}I_{q\ra b}^{(1)}
+{1\over10}\,\varepsilon_{ab\la k}I_{p\ud{a}}^{(6)}I_{q\ra b}
-{1\over2}\,\varepsilon_{ab\la k}I_{p\ud{a}}^{(4)}I_{q\ra b}^{(2)}
-2\,J_{\la k}I_{pq\ra}^{(5)}\right)
\right.\right.\nonumber\\&&\left.\left.
+{4\over63}\,J_{jk}^{(3)}\left(
-{2\over5}\,\varepsilon_{ab\la i}I_{j\ud{a}}^{(5)}I_{k\ra b}^{(1)}
+{1\over10}\,\varepsilon_{ab\la i}I_{j\ud{a}}^{(6)}I_{k\ra b}
-{1\over2}\,\varepsilon_{ab\la i}I_{j\ud{a}}^{(4)}I_{k\ra b}^{(2)}
-2\,J_{\la i}I_{jk\ra}^{(5)}
\right)
\right]
+{\cal O}\left({1\over c^6}\right)\right\},\nonumber\\ 
\end{eqnarray}
\end{widetext}
%%%%%%%%%%%%%%%%%%%%%%%%%%%%%%%%%%%%%%%%%%%%%%%%%%%%%%%%%%%%%%%%%%%%%%%
where,
\begin{widetext}
\label{eq:M2M3S2}
\begin{align}
\left[W^{(2)}I_{ij}-W^{(1)}I_{ij}^{(1)}\right]^{(3)} &=
\left[2\,W^{(4)}I_{ij}^{(1)}+W^{(5)}I_{ij}-W^{(1)}I_{ij}^{(4)}-2\,W^{(2)}I_{ij}^{(3)}\right]
\label{eq:M2}\,,\\
\left[W^{(2)} I_{ijk}-W^{(1)} I_{ijk}^{(1)}+3\, I_{\la ij}Y_{k\ra}^{(1)}\right]^{(4)} &=
\left[W^{(6)}I_{ijk}
+3\,W^{(5)}\,I_{ijk}^{(1)}
+2\,W^{(4)}\,I_{ijk}^{(2)}
-3\,W^{(2)}\,I_{ijk}^{(4)}
\right.\nonumber \\&\left.
-2\,W^{(3)}\,I_{ijk}^{(3)}
-W^{(1)}I_{ijk}^{(5)}
+3\,I_{\la ij}Y_{k \ra}^{(5)}
+12\,I_{\la ij}^{(1)}Y_{k \ra}^{(4)}
\right.\nonumber \\&\left.
+18\,I_{\la ij}^{(2)}Y_{k \ra}^{(3)}
+12\,I_{\la ij}^{(3)}Y_{k \ra}^{(2)}
+3\,I_{\la ij}^{(4)}Y_{k\ra}^{(1)}
\right]
\label{eq:M3}\,,\\
\left[\varepsilon_{ab\la i}\left(-I_{j\ra b}^{(3)}W_{a}
-2\,I_{j\ra b}Y_{a}^{(2)}+I_{j\ra b}^{(1)}Y_{a}^{(1)}\right)
\right]^{(3)} &=\varepsilon_{ab\la i}
\left(
-I_{j\ra b}^{(6)}W_{a}
-3\,I_{j\ra b}^{(5)}W_{a}^{(1)}
-3\,I_{j\ra b}^{(4)}W_{a}^{(2)}
-I_{j\ra b}^{(3)}W_{a}^{(3)}
\right.\nonumber \\&\left.
-2\,I_{j\ra b}Y_{a}^{(5)}
-5\,I_{j\ra b}^{(1)}Y_{a}^{(4)}
-3\,I_{j\ra b}^{(2)}Y_{a}^{(3)}
+I_{j\ra b}^{(3)}Y_{a}^{(2)}
+I_{j\ra b}^{(4)}Y_{a}^{(1)}
\right)
\label{eq:S2a}\,,\\
\left[3\,J_{\la i}Y_{j\ra}^{(1)}-2\,J_{ij}^{(1)}W^{(1)}\right]^{(3)}&=
\left[
3\,J_{\la i}Y_{j\ra}^{(4)}
-2\,J_{ij}^{(1)}W^{(4)}
-6\,J_{ij}^{(2)}W^{(3)}
-6\,J_{ij}^{(3)}W^{(2)}
\right.\nonumber \\&\left.
-2\,J_{ij}^{(4)}W^{(1)}
\right]\,.
\label{eq:S2b}
\end{align}
\end{widetext}
%%%%%%%%%%%%%%%%%%%%%%%%%%%%%%%%%%%%%%%%%%%%%%%%%%%%%%%%%%%%%%%%%%%%%%%
In the above equations, symmetric trace-free projections of the mulitpole moments are denoted by the angular brackets ($\la\ra$) around their indices and the underlined indices  are excluded while taking the projection.
And the hereditary contribution to LMF at 2.5PN is
%------------------------------------------- %---------------------------------------------- %
%-------------------------------LMF hereditary terms----------------------------------- %
%------------------------------------------- %---------------------------------------------- %	

\begin{widetext}
\begin{eqnarray}
\label{LMFhered}
{\left({\mathcal{F}_{i}}\right)_{\rm hered}}&=&
\frac{4\,G^2\,M}{63\,c^{10}}\,I_{ijk}^{(4)}(U)
\int_{0}^{\infty} d\tau \left[\ln \left({\tau\over
	2\tau_0}\right)+{11\over12}\right]
I^{(5)}_{jk}(U-\tau)
+\frac{4\,G^2\,M}{63\,c^{10}}\,I_{jk}^{(3)}(U)\int_{0}^{\infty} d\tau \left[\ln
\left({\tau\over2\tau_0}\right)+{97\over60}\right]
I^{(6)}_{ijk}(U-\tau)
\nonumber\\&&
+{32\,G^2\,M\over 45\,c^{10}}\,\varepsilon_{ijk}\,
I_{ja}^{(3)}(U)\int_{0}^{\infty} d\tau \left[\ln \left({\tau\over
	2\tau_0}\right)+{7\over6} \right]
J^{(5)}_{ka}(U-\tau)
+{32\,G^2\,M\over 45\,c^{10}}\,\varepsilon_{ijk}\,J_{ka}^{(3)}(U)\int_{0}^{\infty} d \tau
\left[\ln \left({\tau\over 2\tau_0}\right)+{11\over12} \right]
I^{(5)}_{ja}(U-\tau)
\nonumber\\&&
+{G^2\,M\over 567\,c^{12}}\,
I_{ijkl}^{(5)}(U)\int_{0}^{\infty} d\tau \left[\ln \left({\tau\over
	2\tau_0}\right)+{97\over60} \right]
I^{(6)}_{jkl}(U-\tau)
+{G^2\,M\over 567\,c^{12}}\,I_{jkl}^{(4)}(U)\int_{0}^{\infty} d \tau
\left[\ln \left({\tau\over 2\tau_0}\right)+{59\over30} \right]
I^{(7)}_{ijkl}(U-\tau)
\nonumber\\&&
+{G^2\,M\over 63\,c^{12}}\,\varepsilon_{ijk}\,I_{jab}^{(4)}(U)\int_{0}^{\infty} d\tau 
\left[\ln \left({\tau\over
	2\tau_0}\right)+{5\over3} \right]
J^{(6)}_{kab}(U-\tau)
+{G^2\,M\over 63\,c^{12}}\,\varepsilon_{ijk}\,J_{kab}^{(4)}(U)\int_{0}^{\infty} d \tau
\left[\ln \left({\tau\over 2\tau_0}\right)+{97\over60} \right]
I^{(6)}_{jab}(U-\tau)
\nonumber\\&&
+{8\,G^2\,M\over 63\,c^{12}}\,J_{ijk}^{(4)}(U)\int_{0}^{\infty} d\tau 
\left[\ln \left({\tau\over
	2\tau_0}\right)+{7\over6} \right]
J^{(5)}_{jk}(U-\tau)
+{8\,G^2\,M\over 63\,c^{12}}\,J_{jk}^{(3)}(U)\int_{0}^{\infty} d \tau
\left[\ln \left({\tau\over 2\tau_0}\right)+{5\over3} \right]
J^{(6)}_{ijk}(U-\tau)
+{\cal O}\left({1\over c^6}\right).\nonumber\\
\end{eqnarray}
\end{widetext}

\section{Orbital dynamics of the compact binary source}\label{orbital_dynamics_LMF}
In the previous section, we have provided an explicit closed form expression of the far-zone linear momentum flux from a compact binary system in terms of various source multipole moments. 
Here we specialize to the case of a non-spinning compact binary system in quasi-elliptical orbits, with the component masses $m_1$ and $m_2$ with $m_1\geq m_2$, the total mass $m=m_1+m_2$, and the symmetric mass ratio, $\eta=m_1m_2/m^2$. Since the binary constituents are nonspinning, its motion is completely confined in a plane with a relative separation,
\begin{align}\mathbf{x}=\mathbf{x_1}-\mathbf{x_2}=r \,\mathbf{\hat{n}}, \end{align} 
with $r=|{\bf x}|$, $\mathbf{x_1}$ and $\mathbf{x_2}$ are the position vectors of the component masses, and $\mathbf{\hat{n}}$ is the unit vector along
the relative separation vector. In polar coordinates,
\be
{\bf \hat{n}}=\frac{\bf x}{r}={\cos{\phi}}\,{\bf \hat{e}_x}+{\sin{\phi}}\,{\bf \hat{e}_y}\,,
\label{eq:n-phi}
\ee
where $\phi$ is the orbital phase of the binary, and $\bf \hat{e}_x$ and $\bf \hat{e}_y$ are the unit vectors along $x$ and $y$ axes. The relative velocity and acceleration for the system are the following,
\begin{align} \label{ellWF:v-a} \mathbf{v}&={{\rm d}\mathbf{x} \over {\rm
			d}t}=(\dot{r}\cos{\phi}-r\dot{\phi}\sin\phi)\,{\bf \hat{e}_x}+(\dot{r}\sin{\phi}+r\dot{\phi}\cos\phi)\,{\bf \hat{e}_y}\,,
	%\,\,\,\,\,\,{\rm and}\,\,\,\,\, \dot{r}={\bf \hat{n}}\cdot {\bf v},
	\\
	\mathbf{a}&={{\rm d}\mathbf{v} \over {\rm d}t}={{\rm d^2}\mathbf{x} \over {\rm d^2}t}.  \end{align}	
To calculate the 2.5PN accurate LMF, we need the time derivative of the source multipole moments. Hence we need 2.5PN accurate equations of motion for the compact binary system. We use the same from  ref.~\cite{BI03CM, ABIQ07}  in the center of mass frame.
The equation of motion can be used to write down the following expressions in order to obtain the derivatives of the multipole moments,
\begin{align} \label{ellWF:vdot-rdot} \dot{v}&={{\bf a}\cdot{\bf v} \over v},\\
	\ddot{r}&={1 \over r}\left[\left(v^2-\dot{r}^2\right)+{\bf a}\cdot{\bf
		x} \right],
\end{align} 
where the $\dot{r}$ and $\ddot{r}$ denote the first and the second time derivatives of the orbital separation $r$ respectively and we denote the magnitude of the orbital velocity by $v=|{\bf v}|$.

To evaluate the instantaneous and the hereditary contributions to the LMF, we would also  need the explicit expressions for the various multipole moments for compact binaries moving in quasi-elliptical orbits. These are obtained from the long algebraic computations (see ref.~\cite{BIJ02} for details) using MPM-PN formalism~\cite{Bliving}. The expressions are too long to be explicitly quoted here.  Hence we point to the Refs.~\cite{BI04,ABIQ07} for those.

\section{Instantaneous contribution to the Linear momentum flux}\label{inst_r_rdot_v_LMF}

With all the ingredients provided in the previous sections we now compute the instantaneous contribution to the LMF using the source multipole moments. First we calculate all the time derivatives of the source multipole moments using the equation of motion as quoted in Ref.~(\cite{ABIQ07}) at  2.5PN. Next we perform all the contractions in Eq.~(\ref{eq:LMF-inst-IJ}) and the resulting instantaneous linear momentum flux in terms of dynamical variables $(r,\dot{r}, v, {\mathbf{v}},{\mathbf{x}})$ is given by,
\begin{widetext}\begin{align}%
		\label{LMFinst-r-rdot-v}
		\left({\mathcal{F}_{i}}\right)_{\rm inst}&=\frac{64}{105}\frac{G^3 m^4}{r^4 c^7} \sqrt{1-4 \eta } \eta ^2\Bigg[ \Bigg(\dot{r}\Big[
		\frac{3}{2}\frac{ G m}{ r}-\frac{45}{8}\dot{r}^2+\frac{55}{8}v^2\Big]+\frac{1}{c^2}\Bigg(
		\frac{G^2 m^2}{r^2}\dot{r}\Big[-\frac{295}{24}+\frac{1 }{12} \eta\Big]+\frac{G m}{r}\dot{r}
		\Big[ \Big(-\frac{4358}{48}+\frac{239}{12} \eta\Big) v^2
		\nonumber\\&
		%------------------------------------------------------------------------------------------
		+\Big(\frac{12301}{144}-\frac{73}{9} \eta \Big) \dot{r}^2\Big]+v^4\dot{r}\Big[\frac{851}{24}-\frac{779}{24} \eta \Big] -v^2\dot{r}^3\Big[\frac{1417}{12}-\frac{1877}{24} \eta \Big] 
		+\dot{r}^5\Big[\frac{1843}{24}-\frac{259}{6} \eta \Big] \Bigg)
		+\frac{1}{c^4}\Bigg(
		\dot{r}^7\Big[\frac{35}{22}+\frac{123565}{264}\eta
		\nonumber\\&
		%------------------------------------------------------------------------------------------
		-\frac{5495}{33}\eta^2\Big]+ \frac{G^3 m^3}{r^3}\dot{r}\Big[\frac{52781}{1584}+\frac{15773}{264} \eta-\frac{607}{264} \eta^2\Big] 
		+\frac{G^2 m^2}{r^2}\dot{r}\Big[\Big(-\frac{803423}{1584}+\frac{5685}{32}\eta-\frac{14437}{528}\eta^2\Big)\dot{r}^2+\Big(\frac{189413}{396}
		\nonumber\\&
		%------------------------------------------------------------------------------------------
		-\frac{754361}{3168}\eta+\frac{26527}{396}\eta^2
		\Big)v^2\Big] - \frac{G m}{r}\dot{r}\Big[ \Big(\frac{329405}{528}-\frac{102581}{96}\eta+\frac{21603}{88}\eta^2\Big)\dot{r}^4
		-\Big(\frac{292697}{352}-\frac{113233}{72}\eta+\frac{463801}{1056}\eta^2\Big)v^2\dot{r}^2
		\nonumber\\&
		%------------------------------------------------------------------------------------------
		+\Big(\frac{76409}{352}-\frac{458683}{1056}\eta+\frac{178873}{1056}\eta^2\Big)v^4
		\Big]-v^2\dot{r}^4\Big[\frac{5363}{132}+\frac{150719}{132}\eta-497\eta^2
		\Big]+ v^4\dot{r}^2\Big[\frac{1327}{33}+\frac{900359}{1056}\eta-\frac{5240}{11}\eta^2
		\Big]
		\nonumber\\&
		%------------------------------------------------------------------------------------------
		+v^6\Big[\frac{85}{44}-\frac{187945}{1056} \eta +\frac{9371}{66} \eta ^2\Big]\Bigg)
		+\frac{G m}{r c^5}\eta\Bigg(\frac{701}{90}v^6-\frac{51137}{96}v^4\dot{r}^2
		+\frac{41611}{40}v^2\dot{r}^4-\frac{49219}{96}\dot{r}^6-\frac{4}{15}\frac{G^3 m^3}{r^3} 
		\nonumber\\&
		%------------------------------------------------------------------------------------------
		+\frac{G^2m^2}{r^2}\Big[\frac{1237}{90}v^2-\frac{6607}{180}\dot{r}^2\Big]-\frac{Gm}{r}\Big[
		\frac{4261}{120}v^4+\frac{8397}{40}v^2\dot{r}^2-\frac{3778}{15}\dot{r}^4\Big]
		\Bigg)
		\Bigg)\frac{\mathbf{x}_i}{r}
		\nonumber\\&
		%------------------To be written------------------------------------------------------------------------
		-\Bigg(\frac{Gm}{r}+\frac{25}{4}v^2-\frac{19}{4}\dot{r}^2
		+\frac{1}{c^2}\Bigg(\frac{G m}{r}\Big[\Big(\frac{2699}{48}+\frac{5}{24}\eta\Big)\dot{r}^2-
		\Big(\frac{907}{16}-\frac{81}{8}\eta\Big)v^2\Big]-\frac{G^2 m^2}{r^2}\Big(\frac{21}{2}+\frac{17}{18}\eta\Big)
		+v^4\Big[\frac{925}{48}-\frac{259}{12}\eta\Big]
		\nonumber\\&
		%------------------------------------------------------------------------------------------
		-v^2\dot{r}^2\Big[\frac{245}{3}-\frac{1285}{24}\eta\Big]
		+\dot{r}^4\Big[\frac{2663}{48}-\frac{697}{24}\eta\Big]
		\Bigg)+\frac{1}{c^4}\Bigg(
		\frac{G m}{r}\Big[
		v^4\Big(-\frac{47107}{352}+\frac{214813}{1056}\eta-\frac{80173}{1056}\eta^2\Big)
		+v^2\dot{r}^2\Big(\frac{199993}{352}
		\nonumber\\&
		%-----------------------------------------------------------------
		-\frac{163247}{176}\eta+\frac{221851}{1056}\eta^2
		\Big)-\dot{r}^4\Big(\frac{120737}{264}-\frac{694453}{1056}\eta+\frac{42029}{396}\eta^2\Big)
		\Big] + \frac{G^2 m^2}{r^2}\Big[
		v^2\Big(\frac{955835}{4752}-\frac{8047}{144}\eta+\frac{3327}{176}\eta^2\Big)
		\nonumber\\&
		%-----------------------------------------------------------------
		-\dot{r}^2\Big(\frac{381131}{1584}-\frac{62105}{1584}\eta-\frac{10855}{1584}\eta^2\Big)
		\Big]+\frac{G^3 m^3}{r^3}\Big[\frac{236347}{4752}+\frac{1913}{44}\eta-\frac{1103}{396}\eta^2\Big]
		+v^6\Big[\frac{7741}{264}-\frac{54215}{528}\eta+\frac{5741}{66}\eta^2\Big]
		\nonumber\\&
		%-----------------------------------------------------------------
		-v^4\dot{r}^2\Big[\frac{73439}{528}-\frac{38405}{66}\eta+\frac{12765}{44}\eta^2\Big] 
		+v^2\dot{r}^4\Big[\frac{30271}{132}-\frac{39085}{48}\eta+\frac{25607}{88}\eta^2\Big]
		-\dot{r}^6\Big[\frac{61339}{528}-\frac{88925}{264}\eta+\frac{24391}{264}\eta^2\Big]
		\Bigg)
		\nonumber\\&
		%-----------------------------------------------------------------
		-\frac{G m \eta}{r c^5}\dot{r}\Bigg(
		\frac{157787}{480}v^4-\frac{39869}{60}v^2\dot{r}^2+\frac{31913}{96}\dot{r}^4
		+\frac{G m}{r}\Big[\frac{10773}{40}v^2-\frac{99277}{360}\dot{r}^2\Big]+\frac{737}{36}\frac{G^2m^2}{r^2}
		\Bigg)
		\Bigg)\mathbf{v}_i\Bigg]. %{\color{red} [Double~checked]} 
\end{align} \end{widetext}

As a consitency check, we confirm that Eq.~(\ref{LMFinst-r-rdot-v}) agrees with the non-spinning contribution to the LMF quoted in ref.~\cite{Racine2008}. One may notice here that the component of the linear momentum flux along the radial direction (i.e. the term associated to the radial direction, $\mathbf{x}$) depends on $\dot{r}$ and hence in case of quasi-circular orbit contributions from these terms are 0 and the emission of linear momentum is along the direction of the relative velocity vector, $\mathbf{v}$. Although this is true only upto relative 2PN order. 

The above expression for linear momentum flux is given in terms of generic dynamical variables $r, \dot{r}$ and $v$. While specializing to the case of quasi-elliptical orbits, it is convenient to express these dynamical variables in terms of the parameters associated with quasi-elliptical orbits, namely the generalized quasi-Keplerian representation (QKR) of the orbital dynamics. 
One needs 2.5PN QKR to compute the 2.5PN  LMF in terms of the orbital parameters. In the next section we briefly start with the description of the parametrization of Keplerian orbits followed by its PN generalization, the quasi-Keplerian (QK) representation.

\section{Keplerian and Quasi-Keplerian parametrization}\label{KR}
The Keplerian parametrization for the Newtonian motion of a compact binary system is widely used in describing celestial mechanics. In polar coordinates and in the center of mass frame, the parametrization is given by,
	\bse
\label{eq:newteqnsQKR}
\begin{align}
	\label{eq:rnewt}
	&r_{\rm N} =a_N(1-e_N \cos u) \, , \\
	\label{eq:phinewt}
	&\phi_{\rm N} = V_N
	\, , \\
	\label{eq:keplereqnewt}
	&l_{\rm N} =n(t-t_0)= u- e \sin u \, , \\
	\label{eq:veqnewt}
	&v= V_{\rm N}(u) \equiv 2 \arctan \left[ \left( \frac{1+e}{1-e}\right)^2 \tan\Big(\frac{u}{2}\Big) \right] \;,
\end{align}
\ese
where subscript N denotes the Newtonian quantities. $r_N$ and $\phi_N$ together define the relative separation vector, $\mathbf{r_N}=r_N(\cos\phi_N,\sin\phi_N,0)$. The semi-major axis of the orbit is $a_N$ with an eccentricity $e_N$. Both of these can be written in terms of the conserved orbital energy and angular momentum which completely define the orbits. Here $u,v, l$ are the eccentric, true, and mean anomalies and  $n$ is the mean motion, $n=2\pi/P$, where P is the radial orbital (periastron to periastron) period.

Having discussed the Keplerian representation (KR), we now describe the PN extension of the KR, the quasi-Keplerian representation at 2.5PN. In 1985, Damour and Deruelle generalized this parametrization up to 1PN \cite{DD85} and proposed a ``Keplerian like parametrization". Later in refs.~\cite{DS88,SW93,Wex95} the 2PN extension of the parametric solution has been quoted. 3PN extension of the same is discussed in ref.~\cite{MGS04}.
3PN QK parametrization is obtained considering 3PN conservative contributions to the binary motion and it admits very similar expressions as the Keplerian one but with more complex structure. In order to obtain 2.5PN accurate LMF, it is sufficient to use 2PN accurate QKR of the orbital motion. Hence, in this section we only describe the 2PN QKR of the conservative dynamics as an extension of Eqs.~(\ref{eq:newteqnsQKR}), 
\begin{widetext}\bse
\label{eq:quasikeeqns}
\begin{align}
	\label{eq:req}
	& r =a_r(1-e_r \cos u) \, , \\
	\label{eq:phi}
	&\phi = \lambda + W(l;n,e_t) \, , \\
	\label{eq:lambda}
	&\lambda = (1+k) n (t-t_0)+c_{\lambda} \, , \\ 
	\label{eq:W}
	&W(l;n,e_t) = ( 1 + k ) ( v - l ) 
	+ \frac{ f_{4\phi} }{c^4} \sin 2 v
	+ \frac{ g_{4\phi} }{c^4}  \sin 3 v
	\, , \\
	\label{eq:keplereq}
	&l = n(t-t_0)+c_l = u - e_t \sin u
	+ \frac{g_{4t}}{c^4} (v - u)
	+ \frac{f_{4t}}{c^4}  \sin v
	\, , \\
	\label{eq:veqn}
	&v = V(u) \equiv 2 \arctan \left[ \left( \frac{1+e_{\phi}}{1-e_{\phi}}\right)^2 \tan\Big(\frac{u}{2}\Big) \right] \, . 
\end{align}
\ese\end{widetext}
The expressions of the functions $f_{4t}$, $g_{4t}$, $f_{4\phi}$, $f_{6\phi}$ and
$g_{4\phi}$ are given in ref.~\cite{MGS04}. $a_r$ is some 2PN equivalent ``semi-major axis". Unlike the KR, in QKR, there appear three eccentricities $e_r, e_t$ and $e_\phi$, instead of one to completely parameterize the motion. These eccentricities can also be written in terms of the 2PN conserved energy and the angular momentum. In the literature, it has been found to be convenient to use only $e_t$ and the mean motion $n$ as the constants of motion and express all the dynamical variables in terms of these two~\cite{Gopu2006}. Additionally one uses a combination of total mass and $n$ given by $\zeta=Gmn/c^3$, as a PN expansion parameter. However, it is equivalent to use $x$ ( $\sim\Big(\frac{G m \omega}{c^3}\Big)^{2/3}$ with $\omega$ being the orbital frequency) and $e_t$ instead of $\zeta$ and $e_t$. Since the convenient choice for a binary moving in quasi-circular orbit would be to use $x$ as the PN expansion parameter, we stick to the variable $x$ to express all our quantities here. 
One may notice that the PN expansion of LMF, Eq.~(\ref{LMFinst-r-rdot-v}) are expressed as a series in $1/c$. One can easily use the relation between $n, c, \zeta$ and $x$ in order to get the correct expression of the concerned quantity at every PN order.
For the convenience of the readers, we provide the explicit expression, used in our calculation, for $\zeta$ in terms of $x,e_t$ below,
\begin{widetext} \begin{align} 
	\label{ellWF:xi2x} 
	\zeta &= {x^{3/2}
		\over(1-e_t^2)^3}\Bigg\{1-3 e_t^2+3 e_t^4-e_t^6+x \Big(-3+6 e_t^2-3
	e_t^4\Big)+x^2 \Big[-\frac{9}{2}+7 \eta +\Big(-\frac{33}{4}-\frac{\eta
	}{2}\Big) e_t^2+\Big(\frac{51}{4}-\frac{13 \eta }{2}\Big)
	e_t^4\Big] \Bigg\}\,.     
	\end{align} \end{widetext}

\section{Instantaneous LMF for compact binaries in terms of quasi-Keplerian parameters in the small eccentricity limit}\label{LMF_inst_et_phi}
We have provided all the necessary ingredients to compute LMF from a compact binary moving in quasi-elliptical orbit in terms of its orbital elements. As a next step, we re-express the instantaneous contribution to the LMF in terms of QK parameters. To be precise, we use Eqs.~(\ref{eq:quasikeeqns}) to re-express Eq.~(\ref{LMFinst-r-rdot-v}) in terms of $\{x, e_t,u,~\&~\phi\}$. For the instantaneous contributions we do not assume the orbital eccentricity to be small and quote the complete closed form expression valid for arbitrary values of eccentricity.
We find it convenient to express the quantities in terms of $\phi$ and $u$ both which helps to obtain the circular orbit limits quite straightforwardly. Finally, we quote the instantaneous contribution to the LMF emitted by a non-spinning compact binary system in a quasi-elliptical orbit. 

\begin{widetext}
\begin{gather}
\label{lmfinst}
\mathcal{F}_{\rm inst}
=
\frac{464}{105} \frac{c^4 }{G}\sqrt{1-4 \eta } \eta ^2 x ^{11/2}	{1\over (1-e_t^2)^2(1-e_t\cos u)^{11}}
\begin{bmatrix}
\mathcal{F}_{\cos\phi}^{\rm inst} & \mathcal{F}_{\sin\phi}^{\rm inst} 
\end{bmatrix}
\begin{bmatrix}
\cos \phi & \sin \phi\\
\sin \phi & -\cos \phi
\end{bmatrix}
\begin{bmatrix}
{\bf \hat{e}_x} \\ {\bf \hat{e}_y}
\end{bmatrix} %{\color{red} Double~checked}
\end{gather}

\begin{align}
	\mathcal{F}_{\cos\phi}^{\rm inst}=\mathcal{F}_{\cos\phi;N}^{\rm inst}+x~\mathcal{F}_{\cos\phi;1PN}^{\rm inst}+x^2~\mathcal{F}_{\cos\phi;2PN}^{\rm inst}+x^{5/2}~\mathcal{F}_{\cos\phi;2.5PN}^{\rm inst}
\end{align}
Where, the explicit contributions at different PN orders are the following,

\begin{align}
\mathcal{F}_{\cos\phi;N}^{\rm inst}= &{e_t\over 58} (1-e_t^2)^2(1-e_t\cos u)^4 (
9-6e_t^2-4e_t\cos u+e_t^2\cos 2u)\sin u\\
%----------------------------------------------------------------------------------
\mathcal{F}_{\cos\phi;1PN}^{\rm inst}=&-{e_t\over 4176}(1-e_t^2)(1-e_t\cos u)^2\sin u
\Big[52\Big(183+29\eta\Big)
+e_t^2\Big(7396-13190\eta-7e_t^2[2986-2727\eta]+e_t^4[10920-7407\eta]\Big)
\nonumber\\ &
%----------------------------------------------------------------------------------
-4e_t\cos u\Big(6948-1754\eta-e_t^2[7369-2586\eta]+e_t^4[3193-832\eta]\Big)
+2e_t^2\cos 2u\Big(1804-95\eta+4e_t^2[404-67\eta]
\nonumber\\ &
%----------------------------------------------------------------------------------
-e_t^4[648-363\eta]\Big) -e_t^3\Big(4\cos [3u][145+2\eta+e_t^2(251-2\eta)]-3 e_t\cos [4u][46+\eta+e_t^2(20-\eta)]\Big)
\Big]
\\
%----------------------------------------------------------------------------------
\mathcal{F}_{\cos\phi;2PN}^{\rm inst}=& -\Bigg(
e_t\Bigg[-{2473\over 5742}-{483137\over 45936}\eta-{1645\over 12528}\eta^2\Bigg]
+e_t^2\Bigg[{1537571\over 34452}+{462007\over 45936}\eta-\frac{481315}{137808}\eta^2\Bigg]\cos u
+e_t^3\Bigg[-{2106595\over 22968}
\nonumber\\ &
%----------------------------------------------------------------------------------
+{15373855\over 275616}\eta+{26021\over 9504}
-\Bigg({233005\over5742}-{4004795\over 275616}\eta+{1027127\over 275616}\eta^2\Bigg)\cos[2u]\Bigg]
+e_t^4\Bigg[\Bigg({1327349\over 34452}-{2556193\over 22968}\eta
\nonumber\\ &
%----------------------------------------------------------------------------------
+{929135\over 34452}\eta^2\Bigg)\cos u 
+\Bigg({120299\over 11484}-{207\over 638}\eta+{8803\over 68904}\eta^2\Bigg)\cos[3u]\Bigg]
+e_t^5\Bigg[\frac{334223}{2088}-\frac{51247793}{1102464}\eta 
-\frac{25252307}{1102464}\eta^2
\nonumber\\ &
%----------------------------------------------------------------------------------
+\Bigg(\frac{1826369}{34452}-\frac{19955}{594}\eta+\frac{1280395}{137808}\eta^2
\Bigg)\cos[2u]
+\Bigg(\frac{38389}{17226}-\frac{603781}{367488}\eta +\frac{61339}{1102464}\eta^2\Bigg)\cos[4u]\Bigg]
+e_t^6\Bigg[\Bigg(-\frac{14490317}{68904}
\nonumber\\ &
%----------------------------------------------------------------------------------
+\frac{65655203}{275616}\eta-\frac{13503565}{275616}\eta^2
\Bigg)\cos u
-\Bigg(\frac{1571803}{45936}-\frac{6006811}{551232}\eta-\frac{37231}{551232}\eta^2
\Bigg)\cos [3u]
-\Bigg(\frac{599}{1392}-\frac{166543}{551232}\eta
\nonumber\\ &
%-----------------------------------------------------------------------------
+\frac{335}{50112}\eta^2\Bigg)\cos [5u]
\Bigg]+e_t^7\Bigg[-\frac{6995893}{183744}-\frac{45896101}{551232}\eta
+\frac{25446325}{551232}\eta^2
+\Bigg({25213079\over 551232}-{22646075\over 2204928}\eta
\nonumber\\ &
%-----------------------------------------------------------------------------	
-{21076241\over 2204928}\eta^2\Bigg)\cos[2u]
+\Bigg({164939\over 551232}-{26933\over 20416}\eta-{46213\over 551232}\eta^2\Bigg)\cos[4u]
+\Bigg({685\over 61248}-{20551\over 734976}\eta+{91\over 734976}\eta^2\Bigg)\cos[6u]
\Bigg]
\nonumber\\ &
%-----------------------------------------------------------------------------	
+e_t^8\Bigg[
\Bigg({137465\over 1584}-{13144529\over 137808}\eta
+{4279205\over 137808}\eta^2\Bigg)\cos u 
+\Bigg({1923539\over 91872}+{393389\over 275616}\eta- {142867\over 275616 }\eta^2\Bigg)\cos [3u]
+\Bigg({281\over 8352}
\nonumber\\ &
%-----------------------------------------------------------------------------
+{89009\over 275616}\eta+{335\over 25056}\eta^2\Bigg)\cos [5u]
\Bigg]
+e_t^9\Bigg[
-{9437\over 1056}+{26338153\over 367488}\eta -{4153147\over 122496}\eta^2 
-\Bigg({58459811\over 1102464}+{1107319\over 367488}\eta
\nonumber\\ &
%-----------------------------------------------------------------------------
-{6780793\over 1102464}\eta^2\Bigg)\cos [2u]
-\Bigg({373913\over 275616}	+{87121\over 367488}\eta-{835\over 1102464}\eta^2\Bigg)\cos [4u]
+\Bigg( {1891\over 122496}-{10073\over 367488}\eta 
\nonumber\\ &
%-----------------------------------------------------------------------------
-{91\over 367488}\eta^2
\Bigg)\cos [6u]\Bigg]
+e_t^{10}\Bigg[
\Bigg({274789\over 11484} +{179717\over 275616}\eta-{1525295\over 275616}\eta^2\Bigg)\cos u
-\Bigg({7345\over 4176}+{146081\over 551232}
-{16189\over 50112}\eta^2\Bigg)\cos [3u]
\nonumber\\ &
%-----------------------------------------------------------------------------
+\Bigg({617\over 4176}+{8275\over 551232}\eta-{335\over 50112}\eta^2\Bigg)\cos [5u]
\Bigg]
+e_t^{11}\Bigg[
-{173407\over 15312} -{543457\over 45936}\eta 
+{184673\over 22968}\eta^2
+\Bigg({3565\over 696} +{4368883\over 734976}\eta
\nonumber\\ &
%-----------------------------------------------------------------------------
-{1584883\over 734976}\eta^2\Bigg)\cos[2u]
+\Bigg({3031\over 15312}-{29509\over 91872}\eta +{2521\over 91872}\eta^2\Bigg)\cos[4u]
-\Bigg({15\over 2552}
-{1493\over 734976}\eta-{91\over 734976}\eta^2\Bigg)\cos[6u]
\Bigg]
\Bigg)\sin u
\nonumber\\ &
%-----------------------------------------------------------------------------
+{1\over 232}\sqrt{1-e_t^2}e_t(1-e_t^2)(5-2\eta)(1-e_t\cos u)^3\sin u
(252-10 e_t\cos u-e_t^2(256-24\cos[2u])-e_t^3(11\cos u-\cos[3u]))
\\
\mathcal{F}_{\cos\phi;2.5PN}^{\rm inst}= &{1\over 10440}(1-e_t^2)^2\eta
\Bigg[-20508+12700e_t\cos[u]-e_t^2(82035-149499\cos[2u])
-e_t^3(56418\cos u+5878\cos [3u])
\nonumber\\ &
%----------------------------------------------------------------------------------
+e_t^4(235851-271998\cos[2u]-1129\cos[4u])
+e_t^5(46952\cos u+4179\cos[3u]+\cos[5u])
-e_t^6(136804
\nonumber\\ &
%----------------------------------------------------------------------------------	
-124158\cos[2u]-1401\cos[4u]-29\cos[6u])
\Bigg]
\end{align}

\begin{align}
\mathcal{F}_{\sin\phi}^{\rm inst}=\mathcal{F}_{\sin\phi;N}^{\rm inst}+x~\mathcal{F}_{\sin\phi;1PN}^{\rm inst}+x^2~\mathcal{F}_{\sin\phi;2PN}^{\rm inst}+x^{5/2}~\mathcal{F}_{\sin\phi;2.5PN}^{\rm inst}
\end{align}
Where, the explicit contributions at different PN orders are the following, 

\begin{align}
\mathcal{F}_{\sin\phi;N}^{\rm inst}= &{1\over 29} (1-e_t^2)^{5/2}(1-e_t\cos u)^4 \Big(\Big[29-4e_t\cos u-(22+3\cos2u)e_t^2\Big]\cos\phi\Big)\\
%----------------------------------------------------------------------------------
\mathcal{F}_{\sin\phi;1PN}^{\rm inst}=& -{1\over 2088}(1-e_t^2)^{3/2}(1-e_t\cos u)^2 \Big(
10848+4556\eta +e_t^2\Big(39054-27676\eta-e_t^2[49881-30512\eta-3e_t^2(3953-2794\eta)]\Big)
\nonumber\\ &
%----------------------------------------------------------------------------------
-e_t\cos u\Big(49878-10136\eta-e_t^2[26067-6446\eta]-e_t^4[8259-2394\eta]\Big)
+2e_t^2\cos [2u]\Big(6708-1396\eta-e_t^2(3255-1276\eta)
\nonumber\\ &
%----------------------------------------------------------------------------------
-3e_t^4(719-4\eta)\Big) 
-3e_t^3\cos[3u](13+70\eta-e_t^2[589+22\eta])
-3e_t^4\cos[4u](151-4\eta+e_t^2[65-14\eta])
\Big)
\\
%----------------------------------------------------------------------------------
\mathcal{F}_{\sin\phi;2PN}^{\rm inst}=& {1\over 116}(1-e_t^2)(5-2\eta)(1-e_t\cos u)^3 \Bigg(638-1588e_t^2+896e_t^4 +(468-458e_t^2+47e_t^4)e_t\cos u-(130-136e_t^2)e_t^2\cos[2u]
\nonumber\\ &
%----------------------------------------------------------------------------------
-3(4-e_t^2)e_t^3\cos[3u]
\Bigg)
+{1\over 1102464}\sqrt{1-e_t^2}\Bigg(
\Bigg[ -33742320 +31536912\eta+2353616\eta^2
\nonumber\\ &
%-----------------------------------------------------------------------------
+e_t^2(226059048-75675888\eta-33433600\eta^2)
-e_t^4(27954300+173455350\eta-122385830\eta^2)
-e_t^6(334128948
\nonumber\\ &
%-----------------------------------------------------------------------------
-474234540\eta+159091188\eta^2)
+e_t^8(173846220-240441438\eta+79270710\eta^2)
-e_t^{10}(8196120-29456064\eta
\nonumber\\ &
%-----------------------------------------------------------------------------
+12233808\eta^2)
+\Big(
e_t(10048432-85118736\eta+18798496\eta^2)
-e_t^3(526309724-447425412\eta+78251360\eta^2)
\nonumber\\ &
%-----------------------------------------------------------------------------
+e_t^5(685925776-543187848\eta+79109344\eta^2)
-e_t^7(106001292-55794972\eta+9227568\eta^2)
-e_t^9(57233736
\nonumber\\ &
%-----------------------------------------------------------------------------
-53777688\eta+9259920\eta^2)\Big)\cos u
+\Big(
e_t^2(109669752-18730992\eta +602064\eta^2)
-e_t^4(3309896+38646144\eta
\nonumber\\ &
%-----------------------------------------------------------------------------
-13112728\eta^2)
-e_t^6(143996252-103863315\eta+20703521\eta^2)
+e_t^8(26676538+6644574\eta+4561366\eta^2)
\nonumber\\ &
%-----------------------------------------------------------------------------
+e_t^{10}(8117568-2160713\eta+1910583\eta^2)\Big)\cos[2u]
+\Big(
-e_t^3(64308804-12941340\eta-2757537\eta^2)
+e_t^5(66212382
\nonumber\\ &
%-----------------------------------------------------------------------------
-10732134\eta+3410608\eta^2)
-e_t^7(11366016+12462156\eta-46648\eta^2)
+e_t^9(9854478+5904870\eta-628440\eta^2)
\Big)\cos[3u]
\nonumber\\ &
%-----------------------------------------------------------------------------
+\Big(
e_t^4(12816100-2181642\eta+407458\eta^2)
-e_t^6(8141924-486948\eta+579380\eta^2)
-e_t^8(3024980+1254162\eta-122770\eta^2)
\nonumber\\ &
%-----------------------------------------------------------------------------
-e_t^{10}(1400904-195072\eta-94296\eta^2)
\Big)\cos[4u]
+\Big(-e_t^5(1457262-654318\eta+33360\eta^2)
+e_t^7(429228+760464\eta
\nonumber\\ &
%-----------------------------------------------------------------------------
+61176\eta^2)
+e_t^9(897354+34578\eta-51576\eta^2)
\Big)\cos[5u]
+\Big(e_t^6(1332-154947\eta-87\eta^2)+e_t^8(67230-99966\eta+4530\eta^2)
\nonumber\\ &
%-----------------------------------------------------------------------------
-e_t^{10}(48960-37509\eta+879\eta^2)\Big)\cos[6u]\Bigg]\Bigg)
\\
\mathcal{F}_{\sin\phi;2.5PN}^{\rm inst}= &-{1\over 10440}(1-e_t^2)^{5/2}\eta\sin u\Big[
890669 e_t-1329517 e_t^3 +476628 e_t^5 
-e_t^2(446788-390148e_t^2)\cos u
\nonumber\\ &
%----------------------------------------------------------------------------------	
 +e_t^3(24447-2667e_t^2)\cos[2u]-2320e_t^4\cos[3u]-600e_t^5\cos [4u]
\Big]
\end{align}
\end{widetext}

We find that in the circular orbit limit where $e_t\rightarrow 0$, Eq.~(\ref{lmfinst}) agrees with the 2PN LMF expression provided in Eq.~(1) of \cite{BQW05} except the 1.5PN term, which is a hereditary contribution discussed later in sec.~\ref{LMF_herd_et_phi}. We also confirm that in the same limit, our instantaneous contribution at 2.5 PN in Eq.~(\ref{lmfinst}), agrees with Eq.~(3.13) of ref.~\cite{MAI12} provided an additional post-adiabatic contribution (which is given in Eq.~(\ref{LMF-PA})) is added.
Next, we cross check various limiting cases of these results. In the Newtonian limit, Eq.~(\ref{lmfinst}) does not represent Eq.~(2.23) of ref.~\cite{Fitchett83} in its present form. By accurately replacing $u$ with $\phi$, we recover the correct limit.
To be noted here, the replacement of $u$ in terms of $\phi$ at each successive order, leads to aperiodic terms, having linear or quadratic dependence on $\phi$ along with all the periodic terms in $\phi$. Here, one would readily agree that only the periodic terms in $\phi$ contributes to the expression of the recoil for quais-circular orbits, since the other contributions are functions of eccentricity and hence 0. The aperiodic terms in $\phi$ arise due to the fact that in this coordinate system defined by $(a_r, e_t~\&~\phi)$, along with the mass asymmetry that gives rise to the emission of LMF in the first place, there is another asymmetry in the orbital motion. %
%along the y-axis. 
Since the origin of the reference frame is at one of the foci of the elliptical orbits, the velocity at the pericenter is not the same as the velocity at the apocenter. As a result, there is a flux of linear momenta emitted along the preferred (towards y-axis in ref.~\cite{Fitchett83})direction. This effect is discussed at the Newtonian order in Eq.~(2.23) of ref.~\cite{Fitchett83}. In order to avoid these aperiodic terms which give rise to the diverging terms w.r.t time, we choose to keep all the expressions in terms of $x,e_t, u$ and $\phi$. 
Having discussed the instanteneous contributions, we now focus on the hereditary contribution at 1.5PN and 2.5PN in the next section.

\section{Hereditary contribution to the linear momentum flux}\label{LMF_herd_et_phi}

Due to non-linearity of the Einstein's field equations, the time varying source moments couple to themselves and to the others. This give rise to the hereditary contributions which depend on the entire history of the system~\cite{BD92}. The leading order hereditary interaction between the mass quadrupole moment ($I_{ij}$) and mass monopole (M or the ADM mass) appears at relative 1.5PN order. In order to estimate the 2.5PN accurate LMF, we need to calculate the 1.5PN and 2.5PN hereditary contributions to it. The explicit contributions to the hereditary part from various source type multipole moments are quoted in Eq.~(\ref{LMFhered}). 

There are different methods proposed in the literature to compute the hereditary contributions. The first one is a semi-analytical method in the frequency domain, proposed in ref.~\cite{ABIQ07tail}. It is based on the Fourier decomposition of Keplerian motion~\cite{PM63}. The general prescription of this decomposition at arbitrary PN order is discussed in \cite{ABIQ07tail}. The Fourier decompositions of the multipole moments at Newtonian order simply read,
\begin{align}
\label{I-Fourier}
I_L(U)&=\sum_{p=-\infty}^{\infty}\mathcal{I}_L e^{ip\ell},\\
\label{j-Fourier}
J_{L-1}(U)&=\sum_{p=-\infty}^{\infty}\mathcal{J}_{L-1} e^{ip\ell},
\end{align}
with the inverse relation to be
\begin{align}
&\mathcal{I}_L=\frac{1}{2\pi}\int_{0}^{2\pi}d\ell I_L(U)e^{-ip\ell}\label{F-domain-IL}\\
&\mathcal{J}_{L-1}=\frac{1}{2\pi}\int_{0}^{2\pi}d\ell J_{L-1}(U)e^{-ip\ell}.\label{F-domain-JL}
\end{align}
All the Fourier coefficients in Eqs.~(\ref{F-domain-IL})~\&~(\ref{F-domain-JL}) can easily be obtained as combinations of Bessel functions. With the correct normalization factors depicted in Eqs (5.1a)~\&~(5.1b) of ref.~\cite{ABIQ07tail}, these coefficients are quoted in Appendix A of the same. This procedure can very well be adopted in order to compute hereditary contributions. However 
another method is proposed in ref.~\cite{Boetzel_2019} where the hereditary integrals are executed in time domain. We employ this method here in order to obtain the 2.5PN hereditary contribution to LMF, $\left({\mathcal{F}_{i}}\right)_{\rm hered}$. 
The closed form expressions of the 2.5 PN hereditary contributions [see  Eq.~(\ref{LMFhered})], consist of several integrations on the various combinations of the source moments over time starting from the remote past to the current retarded time. All the terms are evaluated following the similar procedure.
In this method, we first obtain every integrals in terms of QK variables ${x, e_t, l, \lambda}$ and then perform the time integrations. 
For example we choose the 1st term in Eq.~(\ref{LMFhered}) and for the reader's convenience we rewrite it here, in Eq.~(\ref{F-domain_term1}). 
\begin{widetext}\begin{align}
	\label{F-domain_term1}
	\Big((\mathcal{F}_{i})_{ \rm hered}\Big)_1&=\frac{4\,G^2\,m}{63\,c^{10}}\,I_{\langle ijk\rangle}^{(4)}(t)
	\int_{0}^{\infty} d\tau \left[\ln \left({\tau\over
		2\tau_0}\right)+{11\over12}\right]
	I^{(5)}_{\langle jk\rangle}(t-\tau)
	\end{align}\end{widetext}
To evaluate this integration, we first perform the contraction, $I_{\langle ijk\rangle}^{(4)}(t)I^{(5)}_{\langle jk\rangle}(t-\tau)$ to rewrite them as a sum over all their non-zero components explicitly in terms of the QK variables. As a result, this integral is expressed as a sum over a few integrals of the standard form
\begin{align}
\int_{0}^{\infty} d\tau  e^{i[ \alpha l(t-\tau) +\beta \lambda(t-\tau)]} \left[\ln \left({\tau\over
	2\tau_0}\right)+{11\over12}\right],
\end{align}
which can be reduced to 
\begin{align}
e^{i[\alpha l(t)+\beta\lambda(t)]}\int_{0}^{\infty} d\tau  e^{-i[ \alpha l(\tau) +\beta \lambda(\tau)]} \left[\ln \left({\tau\over
	2\tau_0}\right)+{11\over12}\right],
\end{align}
with the fact that  if $\ell (t)=n(t-t_0)$ at the current time $t$, then at a retarded time ($t-\tau$), $\ell(t-\tau)$ and $\lambda(t-\tau)$ are simply $(\ell(t)-n\tau)$ and $\lambda(t)-\lambda(\tau)$ respectively where $n$ is the mean motion. The above integrals can be solved using the standard formula given by

\begin{align}\label{time-integral}
\int_{0}^{\infty}d\tau e^{i\sigma\tau}\ln\Big(\frac{\tau}{2r_0}\Big)=-\frac{1}{\sigma}\Big[\frac{\pi}{2}Sign(\sigma)+i\Big(\ln(2|\sigma|r_0)+\gamma_E\Big)\Big],
\end{align}
with $\sigma$ being a combination of $\alpha, \beta$, $\gamma_E$ (the Euler constant) and the function $Sign (\sigma)=\pm 1$. In this whole computation, we restrict ourselves in the small eccentricity limit and provide the results accurate up-to $\mathcal{O}(e_t)$. 
Furthermore, since the effects of radiation reaction on the various variables ${x, e_t, l, \lambda}$ starts appearing at relative 2.5PN order, hence neglected for the computation of these integrals.
All the other integrals are similarly computed to find the complete hereditary contributions at 2.5PN.
The two non-zero components of the hereditary contributions to the linear momentum flux in the small eccentricity limit read as, 
\begin{widetext} 
	
	\begin{gather}
	\label{lmf-herd}
	\mathcal{F}_{\rm hered}
	=
	\frac{464}{105} \frac{c^4 }{G}\sqrt{1-4 \eta } \eta ^2 x ^{7}
	\begin{bmatrix}
	\mathcal{F}_{\cos\phi}^{\rm hered} & \mathcal{F}_{\sin\phi}^{\rm hered} 
	\end{bmatrix}
	\begin{bmatrix}
	\cos \phi & \sin \phi\\
	\sin \phi & -\cos \phi
	\end{bmatrix}
	\begin{bmatrix}
	{\bf \hat{e}_x} \\ {\bf \hat{e}_y}
	\end{bmatrix},
	\end{gather}
	
\begin{align}
\mathcal{F}_{\cos\phi}^{\rm hered}=& x^{3/2} \mathcal{F}_{\cos\phi;1.5PN}^{\rm hered} + x^{5/2} \mathcal{F}_{\cos\phi;2.5PN}^{\rm hered},
\end{align}

\begin{align}
\mathcal{F}_{\cos\phi;1.5PN}^{\rm hered} = & -2\log {n\over \omega_0} 
- e_t\Bigg(
\cos[u]\Bigg[
{11277\over 8410} +{95951\over 841}\log 2 -{463563\over 6728}\log 3 + {523\over 29}\log {n\over \omega_0}
\Bigg]-{39\over 29}\pi\sin[u]
\Bigg) 
-e_t^2\Bigg({48657\over 16820}
\nonumber \\&
%------------------------------------------------------------------------------------
+ {2343\over 58}\log {n\over \omega_0} 
+ {945141\over 1682}\log 2 -{4582737\over 13456}\log 3
-{2767\over 464} \pi \sin[2u] 
+\cos[2u]\Bigg[{126007\over 16820}-{463589\over 1682}\log 2 
\nonumber \\&
%------------------------------------------------------------------------------------
-{388557\over 6728}\log 3 +{78125\over 464}\log 5 +{2653\over 58}\log {n\over \omega_0}\Bigg]
\Bigg),
\\
%------------------------------------------------------------------------------------
\mathcal{F}_{\cos\phi;2.5PN}^{\rm hered} = &-{196573\over 50460} +{38917\over 25230}\eta 
-\Big({32835\over 841}-{109740\over 841}\eta\Big)\log 2 
+\Big({77625\over 3364}-{66645\over 841}\eta\Big)\log 3
+\Big({904\over 87}+{1400\over 261}\eta\Big)\log \Big({n\over \omega_0}\Big)
\nonumber\\ &
%------------------------------------------------------------------------------------
+e_t\Bigg(
\cos u \Bigg[-{218401\over 12615}+{783692\over 37845}\eta 
+\Bigg({1307629\over 2523}-{664381\over 7569}\eta\Bigg)\log 2
-\Bigg({8470827\over 53824}+{24589467\over 53824}\eta\Bigg)\log 3
\nonumber\\ &
%------------------------------------------------------------------------------------
-\Bigg({234375\over 1856}-{703125\over 1856}\eta\Bigg)\log 5
+\Bigg({5569\over 87}+{18127\over 261}\eta\Bigg)\log\Big({n\over \omega_0}\Big)
\Bigg] -\Bigg({323099\over 16704}+{25325\over 5568}\eta\Bigg)\pi\sin u
\Bigg)
+e_t^2\Bigg(
-{100693\over 1740}
\nonumber\\ &
%------------------------------------------------------------------------------------
+{17524951\over 302760}\eta
+\Bigg[{39737\over 18}-{4714343\over 841}\eta\Bigg]\log 2
-\Bigg[{1099125\over 3712}-{23785335\over 107648}\eta\Bigg]\log 3
-\Bigg[{26328125\over 33408}-{26328125\over 11136}\eta\Bigg]\log 5
\nonumber\\ &
%------------------------------------------------------------------------------------
+\Bigg[{533\over 6}+{17341\over 87}\eta \Bigg]\log \Big({n\over \omega_0}\Big)
+\cos[2u]\Bigg[
-{196331\over 25230}+{25418317\over 302760}\eta 
-\Bigg({20853451\over 15138}-{44982391\over 7569}\eta\Bigg)\log 2
-\Bigg({75584331\over 107648}
\nonumber\\ &
%------------------------------------------------------------------------------------
+{25176087\over 107648}\eta\Bigg)\log 3
+\Bigg({1203125\over 1152}-{76484375\over 33408}\eta\Bigg)\log 5
+\Bigg({5225\over 58}+{6893\over 29}\eta\Bigg)\log \Big({n\over \omega_0}\Big)
\Bigg]
-\pi\sin[2u]\Bigg[{921673\over 11136}+{1016891\over 33408}\eta\Bigg]
\Bigg).
\end{align}

\begin{align}
\mathcal{F}_{\sin\phi}^{\rm hered}=& x^{3/2} \mathcal{F}_{\sin\phi;1.5PN}^{\rm hered} + x^{5/2} \mathcal{F}_{\sin\phi;2.5PN}^{\rm hered},
\end{align}

Where the contributions are 

\begin{align}
\mathcal{F}_{\sin\phi;1.5PN}^{\rm hered} = & {309\over 58}\pi 
+e_t\Bigg(
{5361\over 116}\pi\cos u
+\Bigg({124698\over 4205}+{95951\over 841}\log 2-{463563\over 6728}\log 3 +{407\over 29}\log \Big({n\over \omega_0}\Big)\Bigg)\sin u
\Bigg)
+e_t^2\Bigg(
{23967\over 232}\pi 
\nonumber \\&
%----------------------------------------------------------------------
+ {51961\over 464}\pi\cos [2u] 
+\Bigg[{219667\over 1682}+{193375\over 1682}\log 2
- {1983609\over 6728}\log 3
+{78125\over 464}\log 5 +{3587\over 58}\log \Big({n\over \omega_0}\Big)
\Bigg]\sin[2u]
\Bigg),
\\
\mathcal{F}_{\sin\phi;2.5PN}^{\rm hered} = & -{2663\over 116}\pi -{2185\over 87}\pi\eta
+e_t\Bigg(
-\Bigg[{628025\over 5568}+{4862903\over 16704}\eta\Bigg]\pi\cos u
+\Bigg[-{5025971\over 25230}-{21195599\over 302760}\eta
-\Bigg({1643716\over 2523}
\nonumber \\ &
%-------------------------------------------------------------------------
-{2639701\over 7569}\eta\Bigg)\log 2
+\Bigg({13725027\over 53824}+{16058907\over 53824}\eta\Bigg)\log 3
+\Bigg({234375\over 1856}-{703125\over 1856}\eta\Bigg)\log 5
-\Bigg({7958\over 87}
\nonumber \\ &
%-------------------------------------------------------------------------
+{1645\over 29}\eta\Bigg)\log \Big({n\over \omega_0}\Big)
\Bigg]\sin u
\Bigg)
+e_t^2\Bigg(-{4328065\over 33408}\pi-{26893253\over 33408}\pi\eta
-\Bigg[{1738843\over 33408}\pi + {30155227\over 33408}\pi \eta\Bigg]\cos [2u]
\nonumber \\ &
%-------------------------------------------------------------------------
+\Bigg[-{16060223\over 25230}-{30465373\over 67280}\eta
-\Bigg({9123587\over 7569}+{5419900\over 7569}\eta\Bigg)\log 2
+\Bigg({162475587\over 107648}-{56210697\over 107648}\eta\Bigg)\log 3
\nonumber \\ &
%-------------------------------------------------------------------------
-\Bigg({14890625\over 33408}-{16484375\over 33408}\eta\Bigg)\log 5
-\Bigg({8280\over 29}+{29548\over 87}\eta\Bigg)\log \Big({n\over \omega_0}\Big)
\Bigg]\sin [2u]
\Bigg),
\end{align}
\end{widetext}
where
\begin{align*}
	\omega_{0} &= \frac{1}{\tau_0}\exp\Bigg[\frac{5921}{1740}+\frac{48}{29}\log 2-\frac{405}{116}\log 3 -\gamma_E\Bigg].
\end{align*}

\section{Post-adiabatic corrections at 2.5PN order}\label{post-adiabatic}

In the section~\ref{KR} we have discussed the conservative quasi-Keplerian description of the binary motion and in the following two sections, obtain the instantaneous and the hereditary contributions to the LMF corresponding to the conservative dynamics where the basic assumption is that the PN parameter x and the eccentricity parameter $e_t$ are the two intrinsic constants of motion. Apart from these two, there are two more extrinsic constants $c_\ell$ and $c_\lambda$ associated to the initial values of the two parameters $\ell$ and $\lambda$ respectively~(see Eq.~\ref{eq:quasikeeqns}). Now in order to obtain LMF at 2.5PN we need to incorporate the effect of radiation reaction on these constants which makes x and $e_t$ to be a time varying function and the equations for the angles $\ell$ and $\lambda$, i.e. Eq.~(\ref{eq:keplereq}) \&~(\ref{eq:lambda}) modifies to 
\begin{align}
l(t)=&\int_{t_0}^t n(t^\prime) dt^\prime +c_l (t)\\
\lambda(t)=&\int_{t_0}^t (1+k(t^\prime))n(t^\prime) dt^\prime +c_\lambda (t)
\end{align}
In order to obtain these complete solutions, one uses PN accurate equation of motion, Eq.~(\ref{ellWF:v-a}) with the correct dissipative contributions(see Ref.~\cite{Boetzel_2019} for a detailed prescription) which were neglected in case of obtaining QK representation given in Eq.~(\ref{eq:quasikeeqns}). Hence, every quantity describing the binary motion have a two scale decomposition: a slow(radiation reaction time scale) secular drift (will be represented by a bar)  and a fast periodic oscillations (orbital timescale)(will be represented by a tilde)
\begin{align}
x(t)&=\bar{x}(t)+\tilde{x}(t),\\
e_t(t)&=\bar{e}_t(t)+\tilde{e}_t(t),\\
c_l(t)&=\bar{c}_l(t)+\tilde{c}_l(t),\\
c_\lambda(t)&=\bar{c}_\lambda(t)+\tilde{c}_\lambda(t) .
\end{align}
We do not show the detailed computation of these quantities here. We simply quote the expressions for the periodic contributions, $\tilde{x},\tilde{e}_t\tilde{c}_l~\& ~\tilde{c}_\lambda$ in terms of their secular counter parts $\bar{x},\bar{e}_t\bar{c}_l~\&~\bar{c}_\lambda$  following ref.~\cite{Boetzel_2019}.

\bse \label{eq: periodic variations}
\begin{align}
\xp(t) =\;& \eta \xb^{7/2} \eb \bigg[ 80 \sin(\lb) + \frac{1436}{15} \eb 
\sin(2\lb) \nonumber\\
&+ \eb^2 \left( \frac{4538}{15} \sin(\lb) + \frac{6022}{45} \sin(3\lb) 
\right) \bigg] \nonumber\\
&+ \bigO(\xb^{9/2}) \,, \\
\ep(t) =\;& -\eta \xb^{5/2} \bigg[ \frac{64}{5} \sin(\lb) + \frac{352}{15} 
\eb \sin(2\lb) \nonumber\\
&+ \eb^2 \left( \frac{1138}{15} \sin(\lb) + \frac{358}{9} \sin(3\lb) 
\right) \bigg] \nonumber\\
&+ \bigO(\xb^{7/2}) \,,\\
\lp(t) =\;& -\eta \xb^{5/2} \bigg[ \frac{64}{5\eb} \cos(\lb) + 
\frac{352}{15} \cos(2\lb) \nonumber\\
&+ \eb \left( \frac{1654}{15} \cos(\lb) + \frac{358}{9} \cos(3\lb) 
\right) \nonumber\\ 
&+ \eb^2 \left( \frac{694}{15} \cos(2\lb) + \frac{1289}{20} \cos(4\lb) 
\right) \bigg] \nonumber\\
&+ \bigO(\xb^{7/2}) \,,\\
\laP (t) =\;& -\eta \xb^{5/2} \bigg[ \frac{296}{3} \eb \cos(\lb) + 
\frac{199}{5} \eb^2 \cos(2\lb) \bigg] \nonumber\\
&+ \bigO(\xb^{7/2}) \,.
\end{align}\ese
As it is evident from the above relations, that the periodic contributions $\tilde{x},\tilde{e}_t\tilde{c}_l~\& ~\tilde{c}_\lambda$ starts at 2.5PN and fully oscillatory in nature and thus correctly describe the post adiabatic corrections to the dynamical variables. We adopt these relations to obtain the 2.5PN post adiabatic corrections to the components of LMF. We use the instanteneous contributions to LMF at Newtonian order (in Eqs.~(\ref{lmfinst})), substitute,
\begin{subequations}\label{eq: phasing subst}
	\begin{align}
	x &\rightarrow \xb + \xp \,, \\
	e_t &\rightarrow \eb + \ep \,, \\
	l &\rightarrow \lb + \lp \,, \\
	\lambda &\rightarrow \lab + \tilde{\lambda} \,,
	\end{align}
\end{subequations}
and replace $\tilde{x},\tilde{e}_t\tilde{c}_l~\& ~\tilde{c}_\lambda$ by their slowly evolving counter parts to obtain the post adiabatic corrections. We find the post adiabatic corrections at 2.5PN to have the following closed-form expressions,
\begin{widetext}
	
	\begin{gather}\label{LMF-PA}
	\mathcal{F}^{\rm PA}
	=
	\frac{464}{105} \frac{c^4 }{G}\sqrt{1-4 \eta } \eta ^3 x ^{8}
	\begin{bmatrix}
	\mathcal{F}_{\cos\phi}^{\rm PA} & \mathcal{F}_{\sin\phi}^{\rm PA} 
	\end{bmatrix}
	\begin{bmatrix}
	\cos \phi & \sin \phi\\
	\sin \phi & -\cos \phi
	\end{bmatrix}
	\begin{bmatrix}
	{\bf \hat{e}_x} \\ {\bf \hat{e}_y}
	\end{bmatrix},
	\end{gather} 
	
	\bse\begin{align}	
	\label{LMFPA}
	\mathcal{F}_{\cos\phi}^{\rm PA}=&{16\over 435}(-54-507 \eb\cos u+\eb^2[2225+20232\cos[2u]]),
	\\
	\mathcal{F}_{\cos\phi}^{\rm PA}=& {2\over 1305}\eb\sin u(315948+620003\eb\cos u).
	\end{align}\ese\end{widetext}

To be noted here that this post adiabatic correction applies only to the Newtonian terms and in rest of the contributions we can safely replace all the variables simply by their secular counter parts.

\section{Total LMF and Log absorption}\label{totalLMF}

In the previous sections, we have presented a closed form expressions for the various contributions to the LMF components in terms of ${\xb, \eb, \ub~\&~\bar{\phi}}$. For simplicity, we first re-express the total LMF components in terms of $\xb, \eb,\lb,\lab$ in stead of ${\xb, \eb, \ub~\&~\bar{\phi}}$. For this purpose we use Eqs.~(25a)~\&~(25b) of Ref.~\cite{Boetzel_2019}. 

Now, as the readers would agree that the hereditary components have some dependence on the arbitrary constant $\tau_0$.  We find that this arbitrary constant can be reabsorbed by a redefinition of the mean anomaly by,

\begin{align}
\xi &= \lb - \frac{3GM}{c^3} \bar{n} \ln \Big( \frac{\xb}{x_0'} \Big) \,,
\end{align}
with $M$ to be the ADM mass and $x_0^\prime=\Big(\frac{G m \omega_0}{c^3}\Big)^{2/3}$. This serves simply as a constant shift to the time coordinate and hence $\xi$ and $l$ follow the same evolution equation, $d\xi/dt=d\lb/dt=\bar{n}$. Similarly, the phase $\lab$ evaluated at the shifted time has the following form,

\begin{align}
\lab_\xi =&\; \lab - \frac{3GM}{c^3} (1 + \bar{k}) \bar{n} \ln \Big( 
\frac{\xb}{x_0'} \Big) \,.
\end{align}
With this shift in the coordinate time, one can also redefine the phase variable as $\psi$. The relation between these variables are given by,
\begin{subequations}
	\begin{align}
	\lb =&\; \xi + 3 \left( \xb^{3/2} - \xb^{5/2} \left( 3 + \frac{\eta}{2} 
	\right) \right) \ln \Big( \frac{\xb}{x_0'} \Big) \,,\\
	\phi =&\; \psi + \bigg( \xb^{3/2} \left( 3 + 6 \eb \cos(\xi) \right) 
	\nonumber\\
	&+ \xb^{5/2} \left( -\frac{3\eta}{2} + 6 \eb (2 - \eta) \cos(\xi) \right) 
	\bigg) \ln \Big( \frac{\xb}{x_0'} \Big) \nonumber\\
	&- 9 \xb^3 \eb \sin(\xi) \ln^2 \Big( \frac{\xb}{x_0'} \Big) \,.
	\end{align}
\end{subequations}
As a result of this redefinition, Eq.~(\ref{lmf-herd}) together with the instantaneous parts in Eqs.~(\ref{lmfinst}) and the post adiabatic corrections in Eq.~(\ref{LMF-PA}) give rise to the total LMF from the system, given in Eq.~(\eqref{total-LMF}) below,
\begin{widetext}
	
	\begin{gather}
	\label{total-LMF}
	\mathcal{F}
	=
	\frac{464}{105} \frac{c^4 }{G}\sqrt{1-4 \eta } \eta ^2 x ^{11/2}
	\begin{bmatrix}
	\mathcal{F}_{\cos[\lambda_\xi]}^{\rm Tot} & \mathcal{F}_{\sin[\lambda_\xi]}^{\rm Tot} 
	\end{bmatrix}
	\begin{bmatrix}
	\cos [\lambda_\xi] & \sin [\lambda_\xi]\\
	\sin[\lambda_\xi] & -\cos [\lambda_\xi]
	\end{bmatrix}
	\begin{bmatrix}
	{\bf \hat{e}_x} \\ {\bf \hat{e}_y}
	\end{bmatrix}
	\end{gather}
	
	\begin{align}
	\mathcal{F}_{\cos[\lambda_\xi]}^{\rm Tot}= & {125\over 58}e_t \sin[\xi]
	-e_t x\Bigg({929\over 348}+{5977\over 1044}\eta\Bigg)\sin[\xi]
	-e_t x^2\Bigg({389\over 1276}-{337585\over 45936}\eta
	-{918679\over 137808}\eta^2\Bigg)\sin[\xi]
	+e_t x^{3/2}\Bigg(12\pi \sin[\xi]
	\nonumber \\ &
	%-------------------------------------------------------------------------------
	-\Bigg[{11277\over 8410}
	+{95951\over 841}\log 2 
	-{463563\over 6728}\log 3\Bigg]\cos[\xi]\Bigg) 
	+x^{5/2}\Bigg(
	{106187\over 50460}-{117774\over 4205}\eta 
	-\Bigg[{32835\over 841}-{109740\over 841}\eta\Bigg]\log 2
	\nonumber \\ &
	%-------------------------------------------------------------------------------
	+\Bigg[{77625\over 3364}-{66645\over 841}\eta\Bigg]\log 3
	+e_t\Bigg[
	\Bigg({464114\over 12615}-{937549\over 2523}\eta 
	+\Bigg[{1307629\over 2523}-{664381\over 7569}\eta\Bigg]\log 2
	-\Bigg[{8470827\over 53824}+{24589467\over 53824}\eta\Bigg]\log 3
	\nonumber \\ &
%-------------------------------------------------------------------------------
	-\Bigg[{234375\over 1856}-{703125\over 1856}\eta\Bigg]\log 5
	\Bigg)\cos[\xi]
	-\Bigg({200123\over 16704}+{334669\over 5568}\eta\Bigg)\pi\sin[\xi]
	\Bigg]
	\Bigg)
	\\
	\mathcal{F}_{\sin[\lambda_\xi]}^{\rm Tot}=& 1+{199\over 29}e_t\cos[\xi]
	-x\Bigg({452\over 87}+{1139\over 522}\eta 
	+e_t\Bigg[{2653\over 116}+{12785\over 522}\eta\Bigg]\cos[\xi]\Bigg)
	+x^{3/2}\Bigg(
	{309\over 58}\pi +e_t\Bigg[{5361\over 116}\pi \cos[\xi]
	+\Bigg[{124698\over 4205}
	\nonumber \\ &
	%-------------------------------------------------------------------------------
	+{95951\over 841}\log 2-{463563\over 6728}\log 3\Bigg]\sin[\xi]
	\Bigg]
	\Bigg)
	+x^2\Bigg(
	-{71345\over 22968}+{36761\over 2088}\eta +{147101\over 68904}\eta^2
	-e_t\Bigg[{3010489\over 34452}-{1623695\over 11484}\eta 
	\nonumber \\ &
	%-------------------------------------------------------------------------------
	-{2793017\over 68904}\eta^2\Bigg]\cos[\xi]
	\Bigg)
	+x^{5/2}\Bigg(
	-{2663\over 116}\pi -{2185\over 87}\pi\eta
	+e_t\Bigg[
	-\Bigg({628025\over 5568}+{4682903\over 16704}\eta\Bigg)\pi\cos[\xi]
	+\Bigg(-{3097214\over 12615}+{4905709\over 12615}\eta
	\nonumber \\ &
   %-------------------------------------------------------------------------------
   -\Bigg({1446706\over 2523}-{664381\over 7569}\eta\Bigg)\log 2
   +\Bigg({11241027\over 53824}+{24589467\over 53824}\eta\Bigg)\log 3
   +\Bigg({234275\over 1856}-{703125\over 1856}\eta\Bigg)\log 5\Bigg)\sin[\xi]
	\Bigg]\Bigg)
	\end{align}
\end{widetext}
To be noted here, although we have computed the total LMF in terms of the secular counter part of all the variables but we have removed the bar while quoting the final expression of the same in Eq.~(\ref{total-LMF}). From now onward we will quote all the quantities in terms of the secular variables but  represent them without the bar.

One may wish to re-express the above equation in terms of the new phase variable $\psi$ and $\xi$ using Eq.~(\ref{Eq:redefined-phase}) below. As a consistency check, in the limit $e_t\rightarrow 0$, above expression reduces to the estimate of LMF from a binary moving in a quasi-circular orbit (see Eq.~(1) of ref.~\cite{BQW05}) once $\lambda_\xi$ is replaced by $\psi$.

\begin{widetext}
	\begin{align}\label{Eq:redefined-phase}
	\lambda_\xi =&\;  \psi -\Bigg[ 2 e_t \sin(\xi) + \frac{5}{4} e_t^2 \sin(2 \xi)+\xb 
	\left( (10- \eta) e_t \sin(\xi) + \left( \frac{31}{4} - \eta \right) 
	e_t^2 \sin(2 \xi) \right) 
	+ x^2 \Big( \frac{1}{12} \left( 624 - 235 \eta + \eta^2 \right) e_t
	\sin(\xi) 
	\nonumber\\&
	+ \frac{1}{24} \left( 969 - 326 \eta + 2 \eta^2 \right) e_t^2 
	\sin(2 \xi) \Big)
	- x^{5/2} \eta \left( \frac{128}{5} + \frac{888}{5} e_t \cos (\xi) + 
	\frac{1}{45} e_t^2 (10728 + 8935 \cos (2 \xi)) \right) \Bigg]\,,
	\end{align}
\end{widetext}

\section{Accumulated LMF}\label{accumulated}
Having discussed the flux of linear momentum, we now compute the total Linear momentum loss in the binary evolution. 
Total loss of linear momentum through the binary evolution over the inspiralling quasi-elliptic orbits can be obtained by integrating the LMF,
\begin{align}
&\frac{d\mathcal{P}_i}{dt}=\mathcal{F}_i,\\
\label{Accumulated-LMF}
&\mathcal{P}_i=\int_{-\infty}^t \mathcal{F}_i dt^\prime.
\end{align}
In order to perform the integration in Eq.~(\ref{Accumulated-LMF}) we follow the prescription for the computation of the memory terms provided in Ref.~\cite{Ebersold_2019} and extend it to the accuracy needed here. As one would immediately identify by replacing Eq.~(\ref{total-LMF}) in Eq.~(\ref{Accumulated-LMF}), that every term is of the form 
\begin{align}\label{osc.Int}
\int_{-\infty}^t dt^\prime x^p e_t^q e^{i(s\lambda_\xi+r\xi)},
\end{align}
where, $p,q,r,s$ are different integers depending on the particular term. From the definition of $\lambda_\xi$, we remind the readers that $\lambda_\xi=(1+k)\xi$ and $\xi=n t$. We quote the final result for integration in Eq.~(\ref{osc.Int}) below,
\begin{widetext}
	\begin{align}
- \frac{\ui}{n(r +s(1+k))}\Bigg(1-\frac{3\ui\chi(T_R)}{8(r+s(1+k))}-\frac{\ui\chi(T_R)}{(r+s(1+k))}\Bigg[-\frac{p}{4}+\frac{19q}{48}\Bigg]\Bigg) x^p\, e^q \, \ue^{\ui (s\lambda_\xi +r \xi)}.
\end{align}
\end{widetext}
We also give the detailed computation in Appendix~\ref{sec:oscintegral}. Furthermore, we observe that there are two types of terms. In the first case, when $r\neq-s$, we identify as the fast oscillatory terms, preserving the usual PN notion in their expansion. On the other hand, when $r=-s$, the terms oscillate on the periastron precession timescale hence slowly oscillating. As a result they are enhance by 1PN. We quote our final accumulated  LMF expressions in Eq.~(\ref{net-LMF}). As evident from Eq.~(\ref{net-LMF}), we observe a relative -1PN and a 0.5PN term appearing in the expression due to the argument presented above. This effect is also seen in case of slow oscillatory memory~\cite{Ebersold_2019}. These terms are not present in case of quasi-circular orbits.

\begin{widetext}
	\begin{gather}
	\label{net-LMF}
	\mathcal{P}= -\frac{464}{105}\sqrt{1-4 \eta } \eta ^2 c m x ^{4}
	\begin{bmatrix}
	\mathcal{P}_{\cos[\lambda_\xi]} & \mathcal{P}_{\sin[\lambda_\xi]}
	\end{bmatrix}
	\begin{bmatrix}
	\cos [\lambda_\xi] & \sin [\lambda_\xi]\\
	\sin[\lambda_\xi] & -\cos [\lambda_\xi]
	\end{bmatrix}
	\begin{bmatrix}
	{\bf \hat{e}_x} \\ {\bf \hat{e}_y}
	\end{bmatrix}
	\end{gather}

\begin{align*}
	\mathcal{P}_{\cos[\lambda_\xi]}&= \frac{91}{116 x}e_t\cos[\xi] +1-e_t\Bigg(\frac{1195}{522}+\frac{301}{232}\eta\Bigg)\cos[\xi] +\sqrt{x} e_t\Bigg(\frac{1323}{232}\pi\cos[\xi]+ \Bigg[\frac{2737}{580}+\frac{364}{45}\eta\Bigg]\sin[\xi]\Bigg)+x\Bigg(-\frac{452}{87} -\frac{1139}{522}\eta \\ \nonumber &
	+e_t\Bigg[\frac{506521}{18729} -\frac{396779}{25056}\eta+\frac{2429}{928}\eta^2\Bigg]\cos[\xi]\Bigg) +x^{3/2} \Bigg(\frac{309}{58}\pi +e_t\Bigg[ \Bigg(-\frac{67663}{6264}-\frac{19729}{783}\eta\Bigg)\pi\cos[\xi] +\Bigg(-\frac{1721569}{50460}+\frac{40249}{841}\log 2 \\ \nonumber &
	-\frac{174069}{6728}\log 3 -\frac{786671}{9396}\eta -\frac{6146}{783}\eta^2\Bigg)\sin[\xi]\Bigg]\Bigg) +x^2\Bigg(-\frac{71345}{22968}+\frac{36761}{2088}\eta + \frac{147101}{68904}\eta +e_t\Bigg[\frac{13818}{145}\pi\eta\sin [\xi] \\ \nonumber &
	+ \Bigg(-\frac{10384631}{30624}+\frac{5356178951}{62013600}\eta +\frac{15582709}{413424}\eta^2 +\frac{17003}{2784}\eta^3\Bigg) \cos[\xi]\Bigg]\Bigg)
	+x^{5/2} \Bigg(
	-\frac{2663\pi}{116}-\frac{2185\pi}{87}\eta + e_t\Bigg[
	\Bigg(\frac{10102429\pi}{33408}
	\\ \nonumber &
	-\frac{80835007\pi}{300672}\eta - \frac{138103\pi}{2349}\eta^2\Bigg)\cos[\xi] + \Bigg(
	\frac{5811883}{25230}+\frac{9725212199}{12261780}\eta-\frac{2842859}{15660}\eta^2 +\frac{603757}{23490}\eta^3 -\Bigg(\frac{1751699}{10092}
	\\ \nonumber &
	-\frac{169934}{7569}\eta
	\Bigg)\log 2 +\Bigg(\frac{2908071}{107648}+\frac{26744067}{107648}\eta\Bigg)\log 3  +\Bigg(\frac{234375}{3712}-\frac{703125}{3712}\eta\Bigg)\log 5
	\Bigg)\sin[\xi]
	\Bigg]
	\Bigg),\\
	\mathcal{P}_{\sin[\lambda_\xi]}&= \frac{91 e_t}{116 x}\sin[\xi]-e_t\Bigg(\frac{7097}{1044}+\frac{301}{232}\eta\Bigg)\sin[\xi]-\sqrt{x} e_t\Bigg(\Bigg[\frac{2737}{580}+\frac{364}{45}\eta\Bigg]\cos [\xi]-\frac{1323}{232}\pi \sin[\xi]\Bigg)
	 \\ \nonumber &
	 +x  e_t \Bigg(\frac{77426}{2349}-\frac{18215}{25056}\eta + \frac{2429}{928}\eta^2\Bigg)+ x^{3/2} e_t \Bigg(\Bigg[\frac{625897}{12615}+\frac{786671}{9396}\eta+\frac{6146}{783}\eta^2+\frac{55702}{841}\log 2-\frac{144747}{3364}\log 3\Bigg]\cos[\xi]
	 \\ \nonumber &
	 -\Bigg[\frac{124997}{3132}+\frac{19729}{783}\eta\Bigg]\pi\sin[\xi]\Bigg)- x^2 e_t \Bigg(\frac{13818}{145}\pi\eta\cos[\xi]+\Bigg[\frac{81690155}{275616} -\frac{1564168363}{31006800}\eta-\frac{11651279}{826848}\eta^2
	 \\\nonumber &
	 -\frac{17003}{2784}\eta^3\Bigg]\sin[\xi]\Bigg)-x^{5/2} \Bigg(\frac{106187}{50460}+\frac{97522}{4205}\eta-\Bigg[\frac{32835}{841}-\frac{109740}{841}\eta\Bigg]\log 2+\Bigg[\frac{77625}{3364}-\frac{66645}{841}\eta\Bigg]\log 3
	 \\ \nonumber &
	 +e_t\Bigg[\Bigg(\frac{35146787}{100920}+\frac{5917848347}{12261780}\eta -\frac{2842859}{15660}\eta^2 +\frac{603757}{23490}\eta^3 + \Bigg[\frac{2029853}{10092}-\frac{494447}{7569}\eta\Bigg]\log 2 - \Bigg[\frac{5678271}{107648}\\ \nonumber &
	 +\frac{22434867}{107648}\eta\Bigg]\log 3 -\Bigg[\frac{234375}{3712}-\frac{703125}{3712}\eta\Bigg]\log 5 \Bigg)\cos[\xi]-\Bigg(\frac{3575993}{11136}-\frac{28032817}{300672}\eta-\frac{138103}{2349}\eta^2\Bigg)\pi \sin[\xi]\Bigg]\Bigg).
\end{align*}
\end{widetext}
As an algebraic test, in the limit $e_t\rightarrow 0$, we recover the correct circular orbit result presented in Ref.~\cite{MAI12} by replacing $\lambda_\xi$with $\psi$ where $\psi$ is the orbital phase in case of compact binaries in quasi-circular orbits.

\section{Conclusion}

In this paper we compute rate of loss of linear momentum in the far-zone of a
nonspinning inspiralling compact binary system in quasi-elliptical orbits. We use QK representation of the orbital variable at 2.5PN. We quote the linear momentum flux accurate upto $\mathcal{O}(e_t)$ at 2.5PN. We also provide a closed form expression for the accumulated linear momentum over the binary evolution. Unlike the linear momentum flux, we observe an additional -1PN and a 0.5PN term appearing in expression for the total linear momentum (see Eq.~(\ref{net-LMF})). This is a very similar effect as seen in computation of the slow oscillatory memory presented in ref~\cite{Ebersold_2019}. 

In most of the previous literature~\cite{MAI12,BQW05}, Eq.~(\ref{Accumulated-LMF}) is used to compute the recoil of the center of mass of the binary where the recoil velocity, $V_i=-\mathcal{P}_i/m$. Although, recently in ref.~\cite{Blanchet:2018yqa}, the authors have argued that the momentum balance equation for the compact binary system will have an additional contribution from the flux associated to the center of mass position. Hence identification of $-\mathcal{P}_i/m$ to the recoil velocity may lead to underestimation of the same. We postpone the calculation of the center of mass flux and the recoil for the future work.

\section{Acknowledgement}
We would like to thank K. G. Arun for priming the author for the topic and for subsequent fruitful discussions. We also acknowledge Chandra Kant Mishra for his valuable suggestions, inputs and cross checking some of the initial computations. Moreover, we thank B. Iyer, G. Faye, M. Favata, C. M. Will for stimulating discussions. Furthermore, we appreciate all the help from the computational facilities at the Institute of Mathematical Sciences, India during the initial phase of this work and the Atlas Computational Cluster team at the Albert Einstein Institute in Hanover for their beyond excellent assistance.

\appendix
\begin{widetext}

\section{Integrals used to compute the accumulated LMF}\label{sec:oscintegral}

We extensively use the method developed for the oscillatory memory integrals in ref.~\cite{Ebersold_2019} to compute the accumulated linear momentum over the binary evolution. However, to achieve the desired accuracy for our calculation, we extend the procedure to one higher order. We provide the complete prescription here. For
convenience we set $G=c=1$ in this appendix. We define the standard integral form (Eq.~(\ref{osc.Int})) that has to be computed as
\begin{align}\label{eq:Jmem}
\mathcal{S}_{\text{osc.}} &= \int_{-\infty}^{T_R}  \ud t\, x^p(t) \, e_t^q(t) \, \ue^{\ui (s \lambda_\xi + r \xi)}\,.
\end{align}
The eccentric orbit is assumed to evolve only with the secular radiation-reaction equations~\cite{Peters&Mathews,Peters1964} given by
\begin{subequations}\label{eq:peters-mathews}
	\begin{align}
	\frac{\ud x}{\ud t} =& \frac{c^3 \eta}{Gm} \frac{x^5}{(1-e_t^2)^{7/2}} \left(\frac{64}{5}
	+ \frac{584}{15} e_t^2 + \frac{74}{15} e_t^4 \right) \,,\label{eq.x-evolution}\\
	\frac{\ud e_t}{\ud t} =& -\frac{c^3 \eta}{Gm} \frac{e_t \, x^4}{(1-e_t^2)^{5/2}} \left(
	\frac{304}{15} + \frac{121}{15} e_t^2 \right) \,,
	\end{align}
\end{subequations}
with $e_t=1$ in
the remote past when $x=0$. We ignore all the astrophysical process such as mass loss during the binary evolution. 

We rewrite the evolution Eq.~(\ref{eq.x-evolution}) accurate upto the leading order in $x$ and keep terms at $\mathcal{O}(e_t^2)$,
\begin{align}\label{eq:dxdt}
\frac{\ud x(t)}{\ud t} = \frac{64 \eta x^5(t)}{5 m}\left[1+ \frac{157}{24} e_t^2(t)\right]\,,
\end{align}
and integrate it over a time interval up to some coalescence time $T_C$, where the orbital
frequency and therefore $x$ tends to infinity:
\begin{align}
\int_{t}^{T_C} \ud t = \int_{x(t)}^{\infty} \frac{\ud x(t)}{(\ud x/\ud t)}\,.
\end{align}
Thereby, we find an explicit relation between the orbital frequency (related to $x$) and time $t$:
\begin{align}\label{eq:tofx}
T_C - t = \frac{5m}{256 \eta}\frac{1}{x^4(t)} \left[1 -\frac{157}{43} e_t^2(t) \right]\,.
\end{align}
%
%We can now invert the $x(e)$ relation derived in~\cref{eq:xofe} to find $e$ as a function of $x$.
Considering only the leading order, we find
\begin{align}\label{eq:eofx}
e_t(t) = e_t(T_R) \left(\frac{x(T_R)}{x(t)} \right)^{19/12}\,.
\end{align}
%
%Using~\cref{eq:tofx,eq:eofx} we get $x$ as an explicit function of $t$:
%
\begin{align}\label{eq:xoft}
x(t) = \frac{1}{4}  \left(\frac{5 m}{\eta (T_C-t)}\right)^{1/4} \left[1 - \frac{157}{172}
\,e_t^2(T_R) \left(\frac{T_C -t}{T_C -T_R}\right)^{19/24} \right]\,.
\end{align}
For our convenience we introduce a new integration variable $y =  \frac{T_R - t}{T_C - T_R}$ and 
the different time-dependent quantities in terms of $y$ and their values at
the current time $T_R$. For $x$ we find
\begin{align}\label{eq:xy}
x(y) = x(T_R)  (1+y)^{-1/4} \left[1 - \frac{157}{172} \,e_t^2(T_R) \left( (1+y)^{19/24} -1\right)
\right]\,,
\end{align}
and for the eccentricity we find
\begin{align}\label{eq:ey}
e_t(y) = e_t(T_R) \left(1+ y\right)^{19/48}\,.
\end{align}
Note that in the remote past as $y\rightarrow \infty$, the eccentricity evolves until it reaches the maximum value of 1. Furthermore, the redefined mean anomaly $\xi(t)$ in terms of $y$ and its
value at the current time is the following,
\begin{align}
\xi(t) = \xi(T_C) - \int^t_{T_C} \ud t' \, n(t') = \xi(T_C) - \frac{1}{m} \int^t_{T_C} \ud t'
\, x^{3/2}(t') = \xi(T_C) - \frac{(T_C - T_R)}{m} \int_{-1}^{y} \ud y' \, x^{3/2}(y') \,.
\end{align}
Now inserting $x(y)$ given
in Eq.~(\ref{eq:xy}) we find,
\begin{align}\label{xittc}
\xi(t) = \xi(T_C) - \frac{8(T_C - T_R) x^{3/2}(T_R)}{5m}(1+y)^{5/8}\left[1 - \frac{471}{11696}
e_t^2(T_R) \left(15(1+y)^{19/24} -34\right)\right] \,,
\end{align}
where $\xi(T_C)$ is the value of $\xi$ at the coalescence. Hence, the mean anomaly at the current time
$T_R$  is given by
\begin{align}\label{eq:lTR}
\xi(T_R) =&\; \xi(T_C) - \frac{8 (T_C - T_R) x^{3/2}(T_R)}{5 m} \left[1 + \frac{8949}{11696}
e_t^2(T_R)\right]\,. 
\end{align}
Now using Eqs.~(\ref{xittc})~\&~(\ref{eq:lTR}) we find $\xi(t)$ in terms of $\xi(T_R)$ and $y$,
\begin{align}\label{eq:xit}
\xi(t) = \xi(T_R) - \frac{8 (T_C - T_R) x^{3/2}(T_R)}{5 m}\left[(1+y)^{5/8}-1\right]\left[1 -
\frac{471}{11696} e_t^2(T_R) \frac{19 -34 (1+y)^{5/8} +15(1+y)^{17/12}}{(1+y)^{5/8}
	-1}\right] \,,
\end{align}
where $x(T_R)$ and $e_t(T_R)$ denote their respective current values of $x$ and $e_t$.

Now we introduce a dimensionless ``adiabatic parameter" $\chi(T_R)$, related to the the inspiral rate at the current retarded time $T_R$ which is defined as the ratio between the
current period and the time difference between the current time and the coalescence,
\begin{align}
\chi(T_R) = \frac{1}{n(T_R)(T_C - T_R)}\,,
\end{align}
where $n(T_R) = x^{3/2}(T_R) / m$ at leading order. We re-write the the adiabatic parameter $\chi(T_R)$ in terms of $x(T_R)$ and $e(T_R)$,
\begin{align}
\chi(T_R) = \frac{256 \eta}{5} x^{5/2}(T_R) \left[1 +\frac{157}{43}e_t^2(T_R)\right] \,.
\end{align}
We express $\xi(t)$ in Eq.~(\ref{eq:xit}) in terms of $\chi(T_R)$,
\begin{align}\label{eq:xiy}
\xi(t) = \xi(T_R) - \frac{8}{5 \chi(T_R)} \left[(1+y)^{5/8} -1\right]\left[1 -
\frac{471}{11696} e_t^2(T_R) \frac{19 -34 (1+y)^{5/8} +15(1+y)^{17/12}}{(1+y)^{5/8}
	-1}\right] \,.
\end{align}
Now using Eqs.~(\ref{eq:xy}),~(\ref{eq:ey})~\&~(\ref{eq:xiy}) , we rewrite Eq.~\ref{eq:Jmem} as an integral
over $y$:
\begin{align}
J_\mem =&\; (T_C-T_R)\int_{0}^{\infty} \ud y \, x^p(y) \, e_t^q(y) \, \ue^{\ui (s\lambda_\xi(y)
	+r \xi(y))}\no
=&\;(T_C-T_R)\,  \ue^{\ui (r+s(1+k)) \xi(T_R)} \int_{0}^{\infty} \ud y \, x^p(y) \, e_t^q(y)
\exp\biggl\{ -\frac{8 \ui (r+s(1+k))}{5 \chi(T_R)} \left[(1+y)^{5/8} -1\right]\no &\times
\left[1 -\frac{471}{11696} e_t^2(T_R) \frac{19 -34 (1+y)^{5/8} + 15(1+y)^{17/12}}{(1+y)^{5/8}
	-1}\right] \biggr\} \,.
\end{align}
Schematically the above integral can be written as,
\begin{align} \label{eq:Jstruc}
J_\mem \sim \int_{0}^{\infty} \ud y \, f(y) \exp \left[ \frac{\ui}{\chi(T_R)}g(y) \right]\,.
\end{align}
We use the technique, integrate by parts, use result from the following type of integral:
\begin{align}
\int \ud y \, \ue^{\ui \sigma g(y)} = -\frac{\ui}{\sigma g'(y)} \ue^{\ui \sigma g(y)} - \frac{g''(y)} {[\sigma^2 g'(y)^3]} \ue^{\ui \sigma g(y)} +
\bigO(g'(y)^{-3}) \,.
\end{align}
This formula is valid as long as $g'(y)$ is sufficiently large. Integrating Eq.~(\ref{eq:Jstruc}) by
parts we get
\begin{align}\label{eq:Jpart}
J_\mem &\sim f(y) \left[ \Bigg(-\frac{\ui \chi(T_R)}{g'(y)}- \frac{\chi(T_R)^2g''(y)} {[g'(y)]^3} \Bigg) \exp \left[
\frac{\ui}{\chi(T_R)}g(y)\right] \right]_{0}^{\infty} + \ui \chi(T_R) \int_{0}^{\infty} \ud
y \, \frac{f'(y)}{g'(y)} \exp \left[ \frac{\ui}{\chi(T_R)}g(y) \right] +\mathcal{O}( \chi(T_R)^3)\,. \\
&\sim  f(y)\left[ \Bigg(-\frac{\ui \chi(T_R)}{g'(y)}- \frac{\chi(T_R)^2g''(y)} {[g'(y)]^3} \Bigg) \exp \left[
\frac{\ui}{\chi(T_R)}g(y)\right] \right]_{0}^{\infty} +\ui \chi(T_R)\times \frac{f'(y)}{g'(y)}\left[ -\frac{\ui \chi(T_R)}{g'(y)} \exp \left[
\frac{\ui}{\chi(T_R)}g(y)\right] \right]_{0}^{\infty} +\mathcal{O}( \chi(T_R)^3)\\
& \sim f(y)\left[\Bigg( \frac{\ui \chi(T_R)}{g'(y)}+\frac{\chi(T_R)^2g''(y)} {[g'(y)]^3}\Bigg) \exp \left[
\frac{\ui}{\chi(T_R)}g(y)\right] \right]_{y=0} +\ui \chi(T_R)\times \frac{f'(y)}{g'(y)}\left[ \frac{\ui \chi(T_R)}{g'(y)} \exp \left[
\frac{\ui}{\chi(T_R)}g(y)\right] \right]_{y=0} +\mathcal{O}( \chi(T_R)^3)\\
& \sim f(y)\left[ \frac{\ui \chi(T_R)}{g'(y)} \exp \left[
\frac{\ui}{\chi(T_R)}g(y)\right]  \Bigg(1-\ui \chi(T_R)\frac{g''(y)}{g'(y)^2}+\ui \chi(T_R)\frac{f'(y)}{f(y)g'(y)}\Bigg)\right]_{y=0}
\end{align}

Now in our case the different functions are the following,
\begin{align}
f'(y)\Bigg|_{y=0}&=\Bigg[px^{p-1}e^q\frac{\ud x}{dy}+qx^pe^{q-1}\frac{\ud e}{dy}\Bigg]_{y=0}=-\frac{p}{4}x^pe_t^q+\frac{19}{48}qx^pe_t^q\\
g'(y)\Bigg|_{y=0}&=-(r+s(1+k)) \\
g''(y)\Bigg|_{y=0}&=\frac{3}{8}(r+s(1+k))
\end{align}

As $y$ approaches infinity in the remote past, the quantity $f(y) = x^p(y) e^q(y)$ and $f^\prime(y)$ both  approach zero. 
since at early times $x\rightarrow 0$ (as the frequency reaches zero) and maximum achievable value for the eccentricity can be 1 in our model. 
While evaluating the terms at $y=0$, we recover $x$ and $e_t$ at the current time
and $g(0) = 0$. The derivative $g'(y)$ in the denominator
evaluated at $y=0$ is effectively 1 multiplied by some constants. 
We keep terms at order $\mathcal{O}(\chi(T_R)^2)$ and the higher-order contributions can be safely ignored. We find the final result for the integral in Eq.~(\ref{eq:Jmem}) as 
\begin{align}
\mathcal{S}_{\text{osc.}} &= -(T_c-T_R) \,x^p\, e^q \, \ue^{\ui (s\lambda_\xi +r \xi)} \frac{\ui \chi(T_R)}{(r
	+s(1+k))}\Bigg(1-\frac{3\ui\chi(T_R)}{8(r+s(1+k))}-\frac{\ui\chi(T_R)}{(r+s(1+k))}\Bigg[-\frac{p}{4}+\frac{19q}{48}\Bigg]\Bigg) \no
&= - \frac{\ui}{n(r +s(1+k))}\Bigg(1-\frac{3\ui\chi(T_R)}{8(r+s(1+k))}-\frac{\ui\chi(T_R)}{(r+s(1+k))}\Bigg[-\frac{p}{4}+\frac{19q}{48}\Bigg]\Bigg) x^p\, e^q \, \ue^{\ui (s\lambda_\xi +r \xi)} \,,
\end{align}
This allows us to compute the oscillatory hereditary integrals in Eq.~(\ref{osc.Int}).

\end{widetext}

\bibliographystyle{apsrev}
\bibliography{ecc_LMF.bbl}
\end{document}